\documentclass[%
twocolumn,   
 showkeys,
unsortedaddress, 
 amsmath,amssymb,
 aps,
 prb,
 floatfix,
]{revtex4-2}

\usepackage{graphicx}
\usepackage{dcolumn}
\usepackage{bm}
\usepackage{color}
\usepackage{braket}
\usepackage{siunitx}
\usepackage{subcaption}
\usepackage{tikz}
\usetikzlibrary{calc}
\usepackage[normalem]{ulem}   

\definecolor{mRed}{RGB}{0,0,0}
\definecolor{mBlue}{RGB}{0,0,0}

\begin{document}

\preprint{APS/123-QED}
\title{Effects of Interactions and Defect Motion on Ramp Reversal Memory in Locally Phase Separated Materials}

\author{Y.~Sun} 
\affiliation{%
Department of Physics and Astronomy, \\Purdue University, West Lafayette, IN 47907, USA
}%
\affiliation{
Purdue Quantum Science and Engineering Institute, West Lafayette, IN 47907, USA
}

\author{M.~Alzate Banguero}
\affiliation{%
Laboratoire de Physique et d'\'Etude des Matériaux, ESPCI Paris, France 
}%
\affiliation{%
Université, CNRS, Sorbonne Université, 75005 Paris, France 
}

\author{P.~Salev}
\affiliation{%
Department of Physics and Astronomy, University of Denver, Denver, Colorado 80208, USA
}%

\author{Ivan~K.~Schuller}
\affiliation{%
Department of Physics and Center for Advanced Nanoscience, University of California-San Diego,
La Jolla, California 92093, USA
}%

\author{L.~Aigouy}
\affiliation{%
Laboratoire de Physique et d'\'Etude des Matériaux, ESPCI Paris, France 
}%
\affiliation{%
Université, CNRS, Sorbonne Université, 75005 Paris, France 
}

\author{A.~Zimmers}
\email{azimmers@espci.fr}
\affiliation{%
Laboratoire de Physique et d'\'Etude des Matériaux, ESPCI Paris, France 
}%
\affiliation{%
Université, CNRS, Sorbonne Université, 75005 Paris, France 
}

\author{E.~W.~Carlson}
\email{ewcarlson@purdue.edu}
\affiliation{%
Department of Physics and Astronomy, \\Purdue University, West Lafayette, IN 47907, USA
}%
\affiliation{
Purdue Quantum Science and Engineering Institute, West Lafayette, IN 47907, USA
}


\date{\today}

\begin{abstract}
The ramp-reversal memory (RRM) effect in metal-insulator transition metal oxides (TMOs), a non-volatile resistance change induced by repeated temperature cycling, has attracted considerable interest in neuromorphic computing and non-volatile memory devices. Our previously introduced defect motion model successfully explained RRM in vanadium dioxide (VO\(_2\)), capturing observed critical temperature shifts and memory accumulation throughout the sample. However, this approach lacked interactions between metallic and insulating domains, whereas the RRM only appears when TMOs are brought into the metal-insulator coexistence regime.
Here, we extend our model by combining the Random Field Ising Model with defect diffusion-segregation, thereby enabling accurate hysteresis modeling while predicting the relationship between RRM and domain interactions. Our simulations demonstrate that maximum RRM occurs when the turnaround temperature approaches the warming branch inflection point, consistent with experimental observations on VO\(_2\). Most significantly, we find that increasing nearest-neighbor interactions enhances the maximum memory effect, thus providing a clear mechanism for optimizing RRM performance.
Since our model employs minimal assumptions, we predict that RRM should be a {\em widespread phenomenon} in materials exhibiting patterned phase coexistence of electronic domains. This work not only advances fundamental understanding of memory behavior in TMOs but also establishes a much-needed theoretical framework for optimizing device applications.
\end{abstract}


\maketitle


\noindent

\section{\label{sec:level1}Introduction}

Neuromorphic computing has recently garnered significant attention as a promising avenue to address the growing energy demands of large language models and other complex artificial intelligence (AI) applications. 
\textcolor{mRed}{
Moreover, local thermal dissipation in conventional architectures may limit packing density and scaling for future AI applications}~\cite{lopez2023ArtificialIntelligenceAdvanced}.
By emulating the structure and function of biological brains, neuromorphic computing offers the potential for energy-efficient computation and a more native hardware platform for AI tasks. One area of considerable interest to facilitate neuromorphic devices is the study of the insulator-metal transition (IMT) in TMOs, which have been explored for their ability to emulate synaptic and neuronal behavior in a manner analogous to biological brains~\cite{Stoliar2017,shi2020}.

\textcolor{mRed}{
In neuromorphic computing, IMT materials are widely used for their unique volatile resistive switching capabilities. 
IMT materials can rapidly switch between high and low resistance states in response to electrical~\cite{salev2023MagnetoresistanceAnomalyElectrical} or thermal~\cite{Feng2023QuantumImagingReconfigurable,zimmersRoleThermalHeating2013} stimuli. These threshold-driven transitions enable the implementation of key neuromorphic components, including spiking neurons that fire when stimulated above a threshold~\cite{boybat2018NeuromorphicComputingMultimemristive} and active signal transmission lines similar to biological axons~\cite{ahmed2023BioinspiredArtificialSynapses}. 
However, volatile switching alone is insufficient for brain-inspired computing, which requires persistent memory to enable learning and adaptation. 
The emerging nonvolatile memory effects in IMT materials, including the ramp reversal memory 
(RRM) observed in TMOs such as VO$_2$, NdNiO$_3$, V$_2$O$_3$, and 1T-TaS$_2$
~\cite{vardiRampReversalMemoryPhaseBoundary2017,anouchiUniversalityMicrostrainOrigin2022,fried2025NewMemoryEffect}
provide a solution while preserving volatile switching characteristics. 
The coexistence of volatile and nonvolatile resistive switching enables IMT materials to serve dual roles as both dynamic switching elements and persistent memory storage in neuromorphic architectures~\cite{khan2024ResistiveSwitchingProperties}.
}

Vanadium dioxide (VO\(_2\)) is a strong candidate for neuromorphic computing hardware, 
\textcolor{mBlue}{as it is potentially able to mimic both neurons and synapses~\cite{IvanK2018neuromorphic,doe-report-schuller-2015}. For neuron behavior, VO\(_2\) has been shown to spike when placed in a simple RC circuit, a phenomenon demonstrated nearly five decades ago~\cite{Rahman_VO2_Oscillator_1977, Fisher_Voltage_Oscillations_VO2_1978}. However, synapse-like connections in VO\(_2\) remain an active research topic, with various approaches being explored, including ion bombardment~\cite{Ghazikhanian2023}, electrical field breakdown~\cite{Cheng2021}, gating~\cite{Anouchi2023}, and thermal coupling~\cite{Velichko2020,Qiu2023,Li2024}. Although each represents a significant step forward, major roadblocks remain: ion bombardment is tunable but not rewritable, and it requires sophisticated material preparation. Electrical field breakdown is easily achievable in any device; however, it is neither reconfigurable nor tunable, due to the uncontrolled stochasticity of filament formation. Gating is slow and highly dependent on sample quality, as it relies on electric-field-induced ion motion toward interfaces. Thermal coupling is not, or is hardly, reconfigurable, as it is primarily determined by the initial lithographic separation of oscillators (0.2 to 20~\unit{\um}~\cite{Velichko2020,Qiu2023,Li2024}). Therefore, new microscopic methods that are tunable, non-stochastic, fast, and easily reprogrammable are necessary 
in order to harness the potential of VO\(_2\)
to truly mimic analog synapse behavior.}

Two distinct properties make VO\(_2\) particularly valuable. First, its thermally induced Mott insulator-metal transition occurs just above room temperature (around 340K)~\cite{Morin1959,liu2024InterestingFunctionalPhase}. This proximity to room temperature, accompanied by a resistivity change of up to five orders of magnitude, enhances its practicality for integration into 
\textcolor{mBlue}{volatile resistive switching devices and neuron-like} applications. 
Second, VO\(_2\) exhibits phase separation during the IMT, a feature that can potentially be controlled and harnessed~\cite{Qazilbash2007}. 
The ability to manipulate the phase-separated state through external factors opens up possibilities for advanced data storage and switching technologies, as well as synaptic device functionality.
Developing a theoretical model to understand and manipulate this phase-separation is crucial to realizing VO\(_2\)'s potential for neuromorphic computing, which in turn provides a pathway toward a revolutionary energy-efficient computer architecture~\cite{Kumar18}.

Recent experimental developments have further heightened this interest in VO\(_2\). In 2017, by applying a ``ramp reversal'' temperature protocol, Vardi \textit{et al.} reported a multi-level, nonvolatile memory effect accompanied by a pronounced ($\sim$20\%) increase in resistivity~\cite{vardiRampReversalMemoryPhaseBoundary2017}. They attributed this memory formation to ``scars''---structural defects analogous to silt deposits at a flood boundary---that raise the local transition temperature and form at phase boundaries during the IMT.
Our earlier \textcolor{mBlue}{experimental observations measuring spatially resolved reflected optical intensity} revealed that the RRM effect occurs throughout the entire sample rather than only at 
phase boundaries, and moreover that it involves both increases and decreases in the local transition temperature (\(T_c\))~\cite{basakSpatiallyDistributedRamp2023}. These results prompted us to develop the defect motion model in which the RRM effect originates from the motion of point defects within VO\(_2\), providing a framework to explain the spatially distributed accumulation of memory. In this model, the redistribution of defect density can either increase or decrease the local \(T_c\)
\textcolor{mBlue}{throughout the bulk of the sample, in addition to producing} 
scar-like structures at phase boundaries, suggesting that our approach may also capture key aspects of the scar model.

\textcolor{mRed}{However, our previous model of RRM did not include interactions between metal and insulator regions, whereas
multiple studies 
have indicated the need for interactions
in order to capture the physics of 
the phase separated regime.}
For example, the presence of avalanches~\cite{sharoniMultipleAvalanchesMetalInsulator2008} in experiments during IMT
provides direct evidence for interactions.
Moreover, both our deep learning classifier~\cite{Basak2023DLHamiltonians} 
and 
the critical exponents reported by us for avalanches~\cite{liuRandomFieldDriven2016}
further indicate
that interactions are necessary in order to explain
the morphology of the domains.
Since our previous defect motion model did not include these interactions, it could not reproduce avalanche phenomena.
These insights motivate the introduction of such interactions into our current model, significantly extending the foundations laid in our earlier work.
\textcolor{mRed}{As we will show in this work, these interactions not only increase the
hysteresis width (which now better matches the hysteresis observed in experimental curves), but they also lead to a marked enhancement of the RRM effect.}

Here, we develop an Interacting Defect Motion Model to describe the thermally driven evolution of the spatial distribution of the memory effect in TMOs. This model builds upon our previous work by incorporating Ising-like nearest-neighbor interactions, enabling us to reproduce the width of the hysteresis loop and to predict the dependence of the maximum RRM effect on the interaction strength. By employing a Correlated Random Field Ising Model (C-RFIM) to simulate the IMT, we hypothesize that the memory effect emerges from the interplay between point defect diffusion and the interactions inherent in the IMT itself. To test this hypothesis, we simulate the metal-insulator coexistence region under a specified temperature protocol designed to replicate the conditions of ramp-reversal experiments. Our simulations reproduce the temperature-dependent behavior of RRM observed experimentally in VO$_2$, thereby providing strong support for our theoretical framework.



\section{Models and Methods}

To develop a more comprehensive framework for the transition behavior in TMO thin films, we have integrated our defect motion model with the C-RFIM. This coupled approach enables us to reproduce key experimental observations while eliminating our previous dependence on experimental \(T_c\) maps as model inputs. By incorporating interactions between neighboring sites, 
our model presented here combining defect diffusion with metal-insulator domain interactions
captures more features of the overall hysteresis behavior as well as the RRM effect. 

\subsection{Correlated Random Field}

\begin{figure*}[bt]
    \centering
    \def\xposition{-10}
    \def\yposition{165}
    
    \begin{picture}(0,0)
    \put(\xposition,\yposition){{\footnotesize (a)}}
    \end{picture}
    \includegraphics[width=0.48\textwidth]{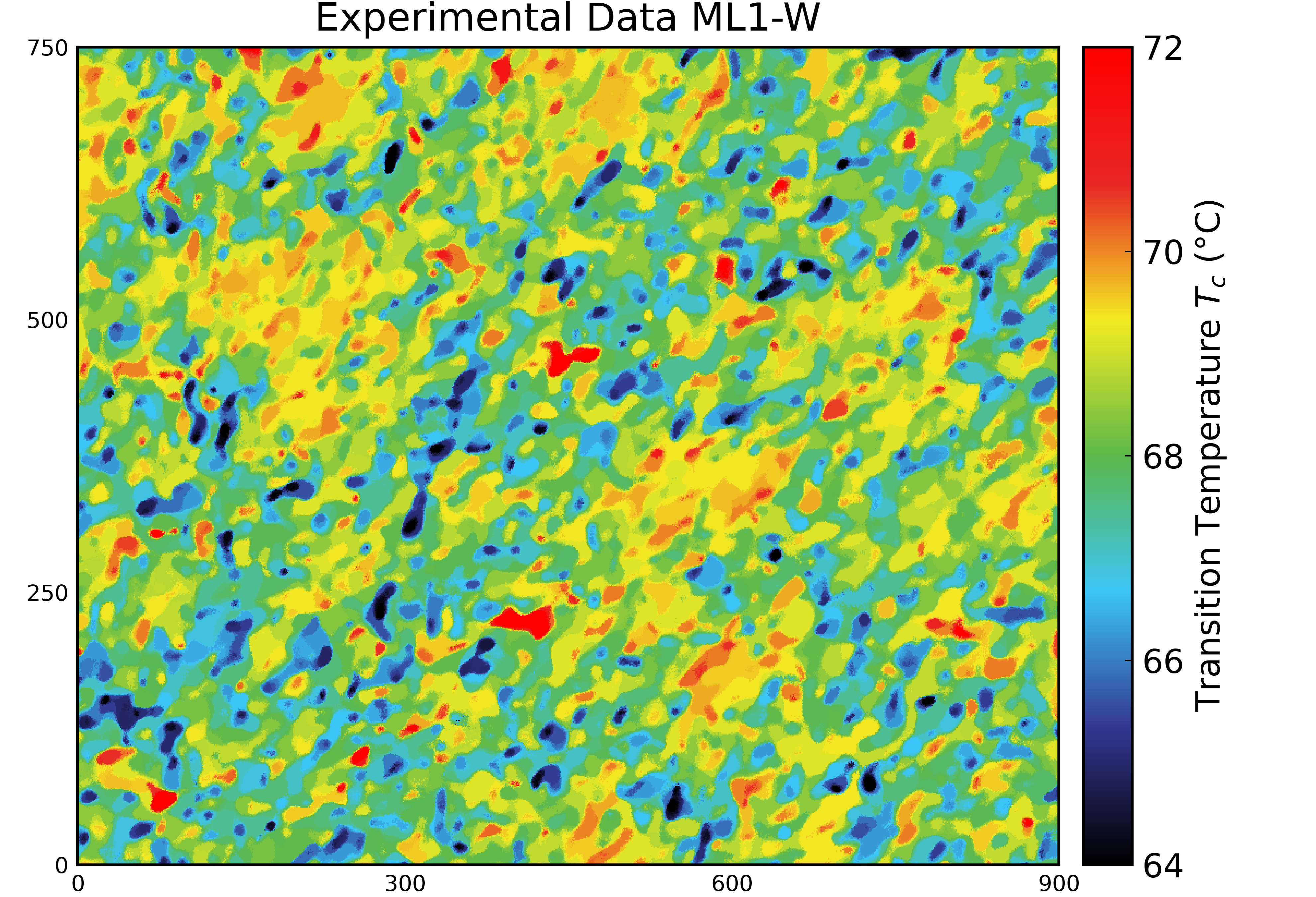}    
    \begin{picture}(0,0)
        \put(\xposition,\yposition){{\footnotesize (b)}}
    \end{picture}
    \includegraphics[width=0.48\textwidth]{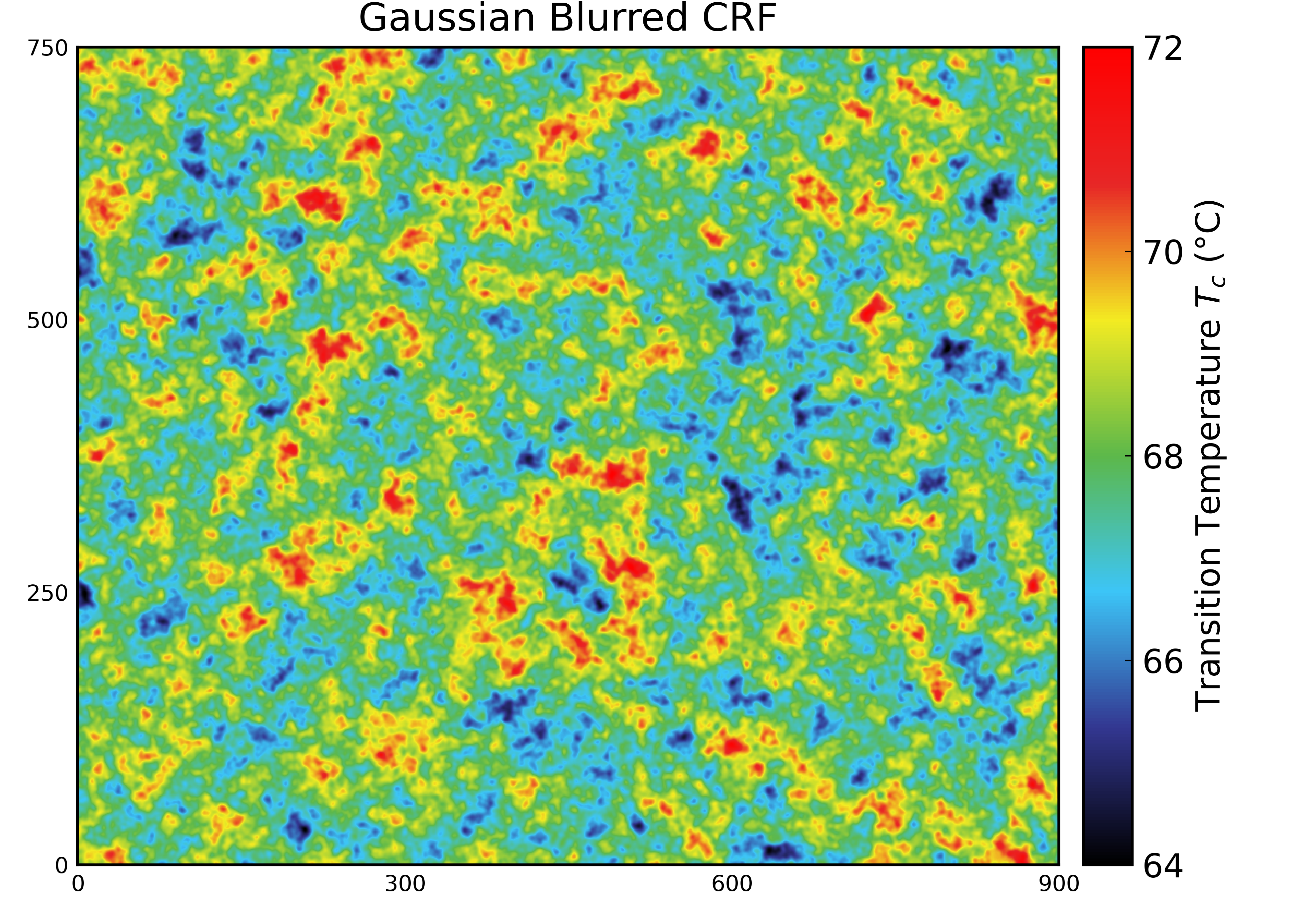}\\

    \caption{
        (a) Experimental \(T_{c}\) map measured during the 
        first warming process.
        (b) Theoretical \(T_{c}\) map using correlated random field (CRF) with Gaussian blurring effect to facilitate comparison
        with the experimental resolution of the \(T_c\) map in Panel (a).
    }
    \label{fig:corRFIM}

\end{figure*}
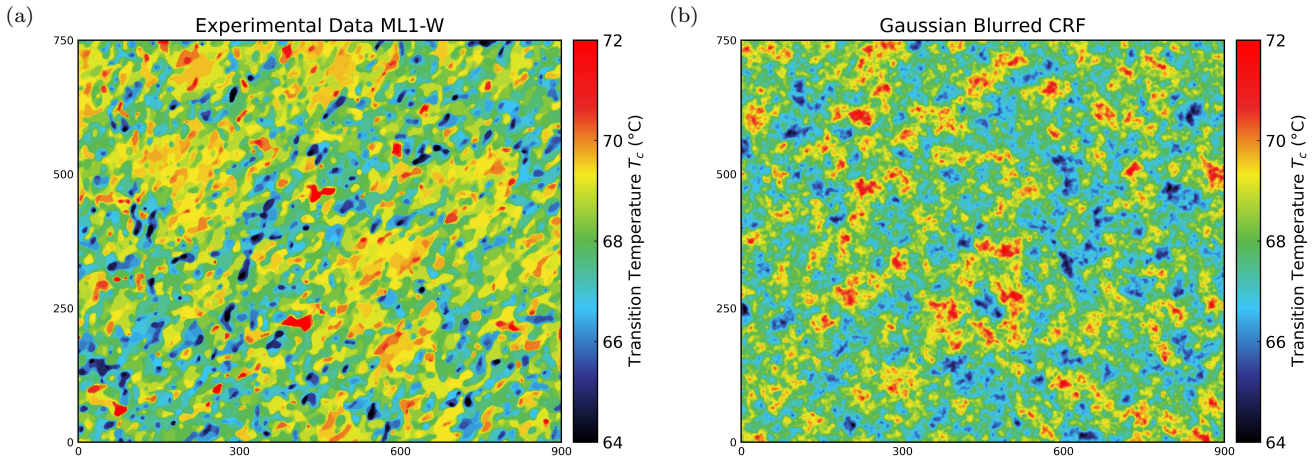

Figure~\ref{fig:corRFIM}(a) shows an example of an experimentally
derived map of the local transition temperature \(T_c\).
Here, we have imaged the metal-insulator phase separation state
on a thin film of VO$_2$
during the warming branch of the initial major hysteresis loop (ML1-W) using high-resolution optical microscopy. 
By slowly heating the sample from a fully insulating state to a fully metallic state, we record the \(T_{c}\) at each pixel (see Figure~\ref{fig:corRFIM}(a)) 
to produce a \(T_{c}\) map through 
the DOMain INtensity Overturn (DOMINO) 
procedure we introduced in~\textcite{basakSpatiallyDistributedRamp2023}. 

In order to create a purely theoretical model, we use initial 
\(T_c\) maps that are generated by statistical means, to have similar
statistics and spatial correlation characteristics to those
observed experimentally.  
The \(T_c\) map Figure~\ref{fig:corRFIM}(a) indicate random field behavior, with regions of
high and low critical temperatures extend over 
a range of
length scales, 
demonstrating inherent spatial correlations
rather than uncorrelated disorder (see the Supplementary Information).
This correlation necessitates
the use of a correlated random field, rather than
an uncorrelated Gaussian random field typically used in
the RFIM. 

Therefore, to create the initial theoretical \(T_c\) map, we first construct an uncorrelated random field with independent identical standard Gaussian distributions. Then we apply Cholesky decomposition (see the Supplementary Information) to transform the uncorrelated random field into a spatially correlated random field. 
Finally, to account for experimental conditions, we apply Gaussian blurring to the correlated random field.
Figure~\ref{fig:corRFIM}(b) shows a \(T_c\) map in theory with Gaussian blurring applied, which reproduces the smoothing effect observed in optical measurements in Figure~\ref{fig:corRFIM}(a) (see the Supplementary Information for detailed comparison).

This initial theoretical \(T_{c}\) map is considered a replacement of the experimental maps with minimal assumptions. By using the initial theoretical \(T_{c}\) map as new model inputs, we eliminate data dependence and enable a more self-consistent theoretical description of RRM.
Note that the RRM temperature protocol (see below) will ultimately modify this initial \(T_c\) map.

\subsection{Random Field Ising Model}

{\color{mBlue}
As discussed in the Introduction, 
avalanches in VO\(_2\)~\cite{basakSpatiallyDistributedRamp2023,sharoniMultipleAvalanchesMetalInsulator2008} and in V$_2$O$_3$~\cite{avalanches-v2o3}
indicate the presence of interactions in the system.
We introduce interactions via a nearest-neighbor Ising interaction,
mapping the two phase metal-insulator system onto a two-dimensional zero-temperature Ising model. In this mapping, metallic and insulating domains correspond to a pseudospin 
at each lattice site
\(\sigma_i = \pm1\), respectively. 
As in Ref.~\cite{liuRandomFieldDriven2016}, 
we map experimental temperature to a uniform external field
that drives the transition from insulator to metal.  
Our previous machine learning classification~\cite{Basak2023DLHamiltonians} indicates that the metal/insulator domain maps of VO\(_2\) belong to the universality class~\cite{Stanley1999UniversalityThreePillars} of the 2D Random Field Ising Model~\cite{Sethna1993Hysteresis}.
To capture the spatial stochasticity observed in experimental \(T_{c}\) maps~\cite{basakSpatiallyDistributedRamp2023},
there must be a random field assigned at each site $i$. 

We thus adopt the 2D Correlated Random Field Ising Model (2D C-RFIM) to model the IMT behavior in VO\(_2\) thin films:
}
\begin{equation}
    H = -J\sum_{\langle i,j \rangle} \sigma_i \sigma_j - \sum_i f\big(T - T_c^{(i)} \big) \sigma_i,
    \label{eqn:Hamiltonian}
\end{equation}
where \( J \geq 0 \) is the interaction strength between neighboring pseudospins, and \( \sigma_i = \pm 1 \) denotes the state of site \( i \), with \( \sigma_i = +1 \) representing metallic and \( \sigma_i = -1 \) representing insulating. The function \( f \) is a monotonic increasing function satisfying \( f(0) = 0 \). The term \( T_{c}^{(i)} \) represents the local critical temperature at site \( i \), generated from the correlated random field. Here, \( f(T - T_{c}^{(i)}) \) acts as an effective external field that biases the local Ising variable \(\sigma_i\) toward a particular state. In the limit of \( J = 0 \), the system becomes non-interacting, and each site undergoes the IMT independently when the temperature \( T \) exceeds its \( T_{c}^{(i)} \).


\subsection{Defect Motion}

To account for the spatial variations in the \(T_{c}\) map observed in our previous work---where \(T_{c}\) increases in some regions and decreases in others as a result of the RRM process, and moreover \(T_c\) changes throughout the entire sample rather than only at scars (see Fig.~S7 in Ref.~\cite{basakSpatiallyDistributedRamp2023})---we proposed that the local \(T_{c}\) change is influenced by the change of defect density \(\rho\) at each site:
\begin{equation}
    \mathrm{\Delta}T_{c}^{(i)} = -\alpha\mathrm{\Delta}\rho_i
    \label{eqn:rho-T}
\end{equation}

The evolution of the defect density is then modeled using the 2D diffusion-segregation equation~\cite{youSimulationTransientIndiffusion1993}:
\begin{equation}
    \frac{\partial \rho(\vec{r})}{\partial t} = \nabla\left[ D\,\Big(\nabla\rho(\vec{r}) - \frac{\rho(\vec{r})}{\rho_{\textnormal{eq}}(\vec{r})}\nabla\rho_{\textnormal{eq}}(\vec{r})\Big) \right],
    \label{eqn:diffusion-segregation}
\end{equation}
where \(D=D(\sigma_{\vec{r}}, T)\) is the diffusion coefficient, and \(\rho_{\textnormal{eq}}\) is the equilibrium density, assigned two distinct values within the metallic and insulating regions. To prevent non-physical behavior near the phase boundary, we apply Gaussian smoothing to the segregation current of defects (See SI).
The additional term involving \(\nabla\rho_{\textnormal{eq}}\) in Eqn.~\ref{eqn:diffusion-segregation}, compared to the well-known diffusion equation, arises from the nonuniform chemical potential near the metal-insulator phase boundaries. This term becomes significant near phase boundaries as indicated in Figure~\ref{fig:lattice} and vanishes deep inside the metallic or insulating regions. The ratio of the equilibrium defect density deep inside two phases \(\rho{_\textnormal{eq}^\textnormal{insulator}}/\rho_\textnormal{eq}^\textnormal{metal}=s\) is known as the ``segregation coefficient''. Other parameters are listed in the Supplementary Information.

\subsection{Dynamics of Defect Motion}

In the presence of both defect motion and Ising-type interactions, the simulation proceeds on a square lattice, as illustrated in Figure~\ref{fig:lattice}. Each lattice site represents a pixel in the experimental images, associating with three local properties that evolve over time: the Ising variable \(\sigma_i\) denoting whether the site is insulating or metallic, the local critical temperature \(T_{c}^{(i)}\), and the local defect density \(\rho_i\). The simulation proceeds through the following steps:

\begin{figure}[h]
    \centering
    \includegraphics[width=0.48\textwidth]{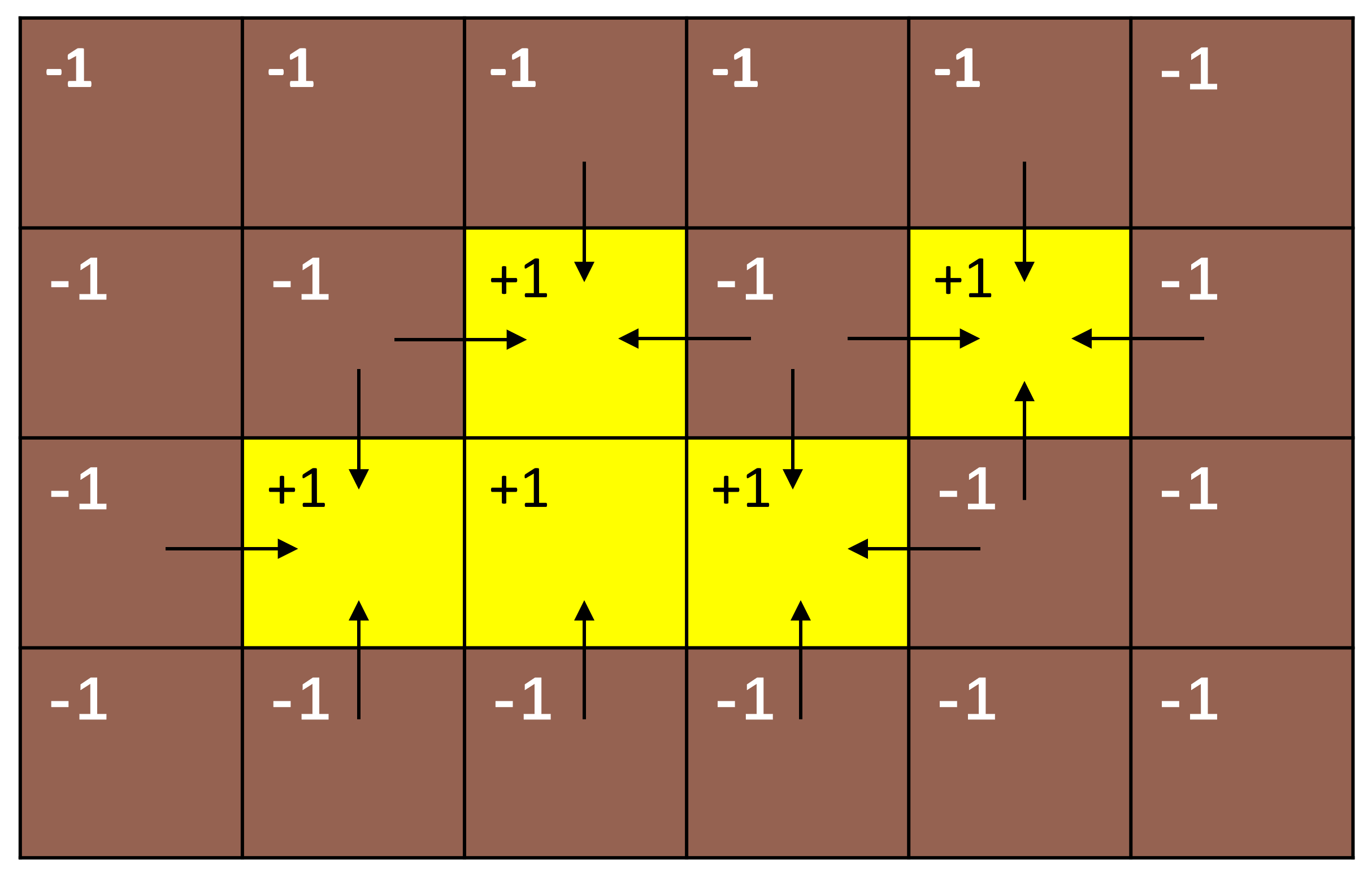}
    \caption{Schematic representation of a portion of the lattice used in the simulation. Arrows indicate the defect current driven by the gradient of chemical potential. Labels with Ising variable \(+1\) and \(-1\) represent metallic (yellow) and insulating (brown) regions, respectively.
    In the mixed phase, defects concentrate preferentially in metallic regions.
    }
    \label{fig:lattice}
\end{figure}

\begin{enumerate}
    \item Calculate the defect distribution using the diffusion-segregation equation [Eq.~\eqref{eqn:diffusion-segregation}].
    \item Compute the critical temperature at each site based on the local defect density using Eq.~\eqref{eqn:rho-T}.
    \item Update the second term of RFIM [Eq.~\eqref{eqn:Hamiltonian}] to reflect the new critical temperatures.
    \item Update the non-equilibrium Ising variable \(\sigma_i\) configuration of the \textcolor{mBlue}{C-}RFIM. Employ the checkerboard update method to iteratively update the pseudospins, accepting new configurations that lower the system's energy until it reaches a local minimum.
    \item Update the 
    equilibrium defect density \(\rho_{\textnormal{eq}}\) and the diffusion coefficients based on the updated pseudospin variables.
\end{enumerate}

These steps collectively constitute one iteration of our simulation process.

\section{Results and Discussion}

In this section, we present our simulation results that model the memory effect observed in VO\(_2\) and discuss the underlying mechanisms.

\subsection{Hysteresis Loop of VO\(_2\)}

\begin{figure*}[ht]
    \centering
    \includegraphics[width=0.9\linewidth]{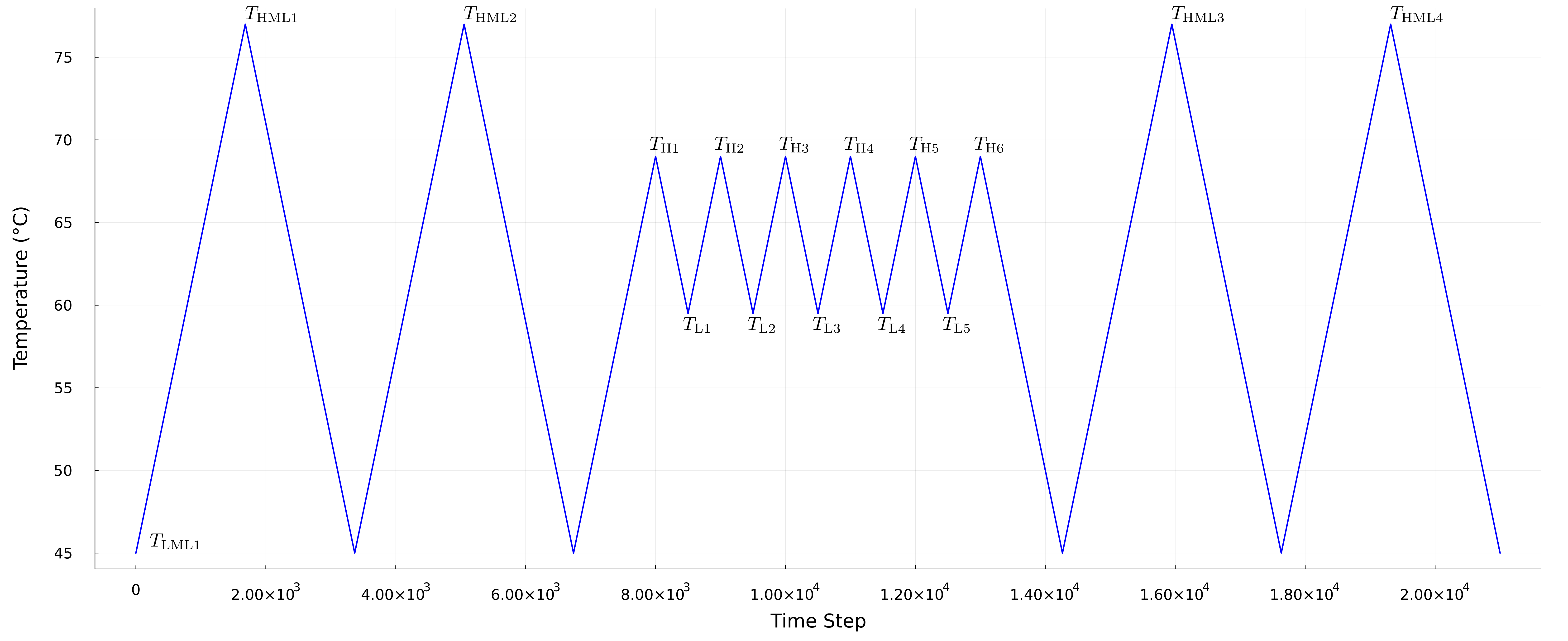}\\
    \caption{(a) Temperature protocol used in our simulation. 
    }
    \label{fig:5-phases}
\end{figure*}

\begin{figure*}[htb]
    \centering
    \begin{picture}(0,0)
        \put(-10,160){{\footnotesize (a)}}
    \end{picture}
    \includegraphics[width=0.85\linewidth]{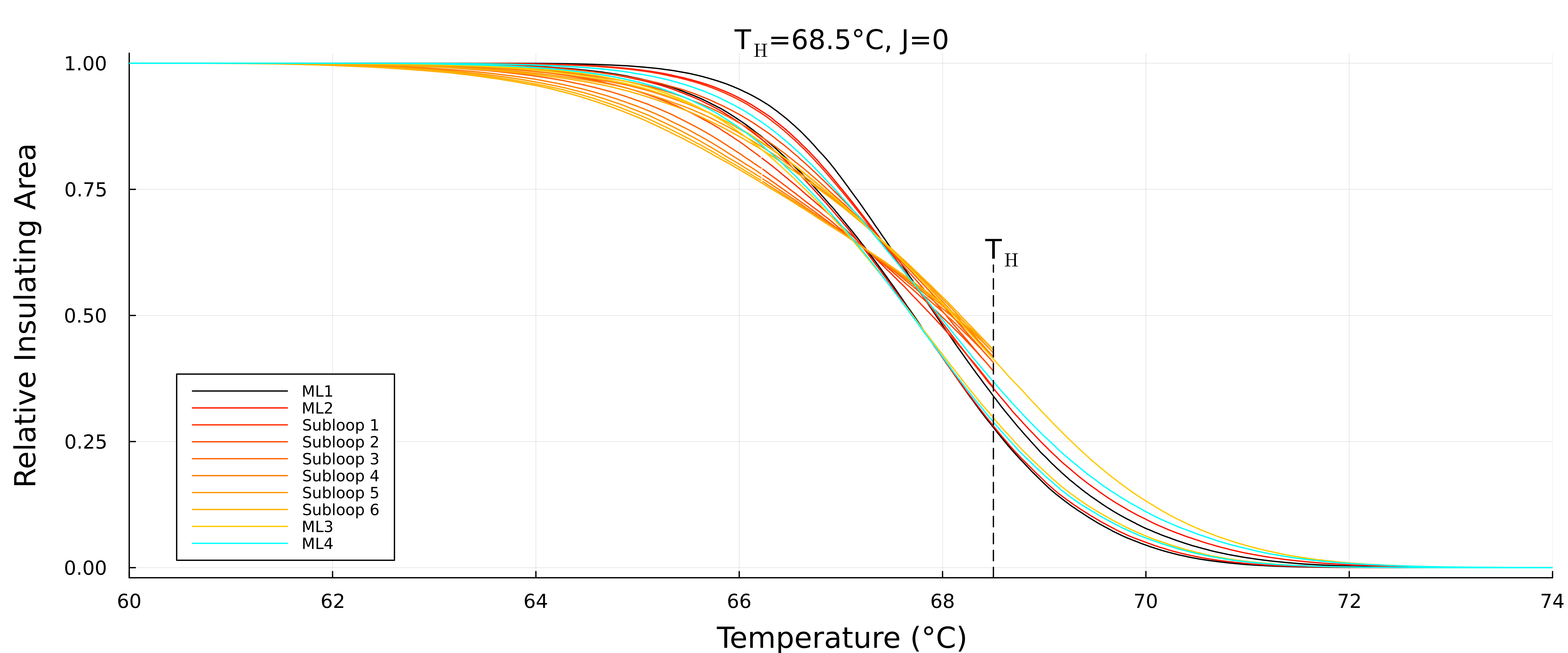}\\
    
    \begin{picture}(0,0)
        \put(-10,160){{\footnotesize (b)}}
    \end{picture}
    \includegraphics[width=0.85\linewidth]{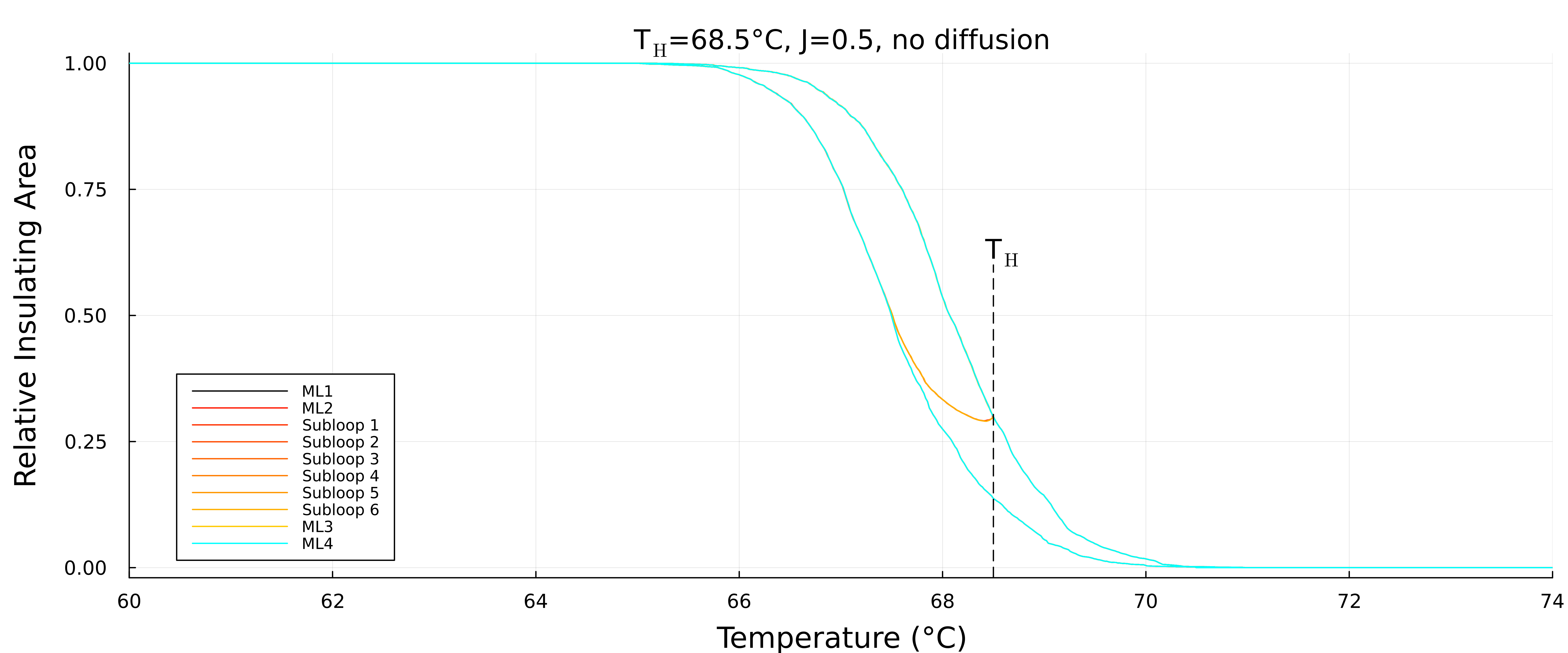}\\

    \begin{picture}(0,0)
        \put(-10,160){{\footnotesize (c)}}
    \end{picture}
    \includegraphics[width=0.85\linewidth]{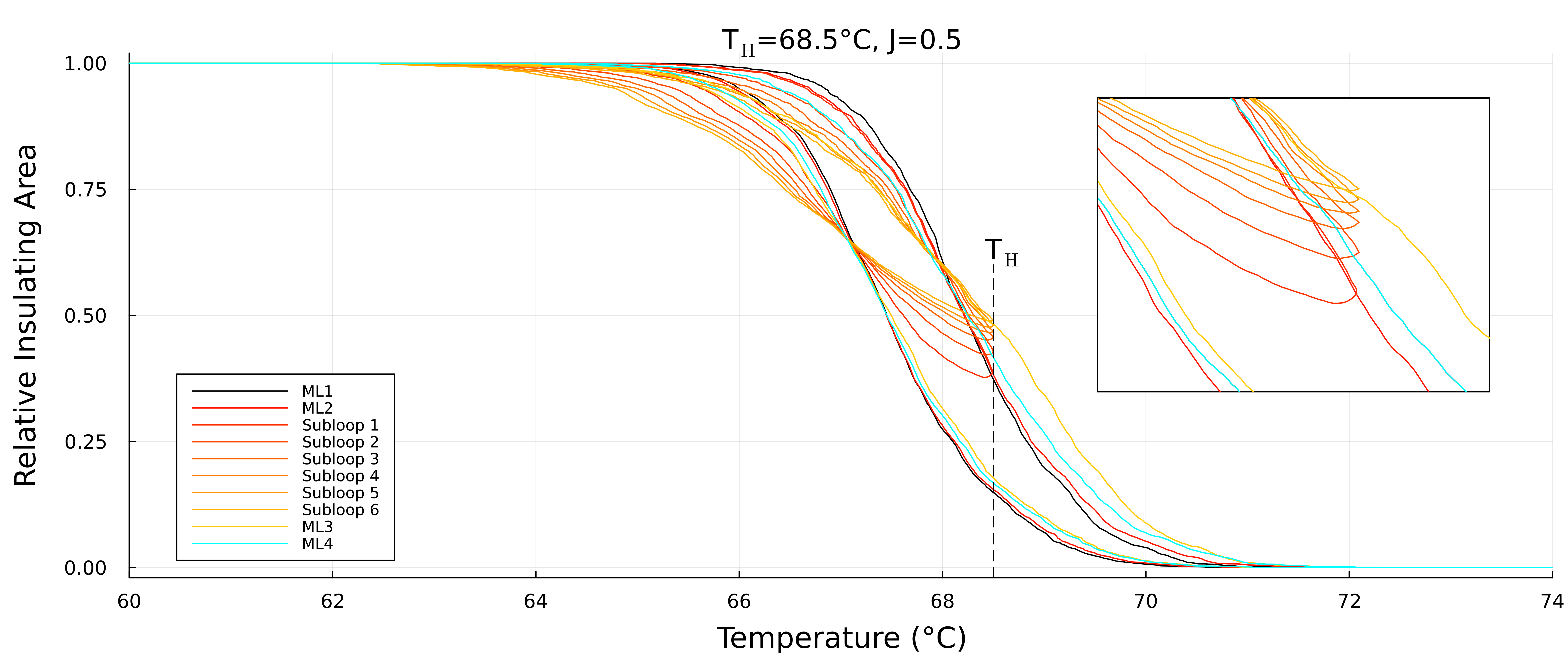}\\
    \caption{Simulated hysteresis loops of the model under different conditions. The black curve represents the initial major hysteresis loop (ML1), while the yellow curve corresponds to the final major loop (ML4) after six subloops. The progression from dark red to orange to yellow indicates the sequence of the loops. (a) Results with defect motion but without interactions (\(J=0\)). (b) Results with interactions (\(J=0.5\)) but without defect motion. Loops stack on top of each other since there is no RRM effect. (c) Results with both defect motion and interactions (\(J=0.5\)) included. \textcolor{mRed}{Subloops have counter-clockwise rotations.}
    }
    \label{fig:loops}
\end{figure*}
To simulate the hysteresis behavior of VO\(_2\), we employed a temperature protocol 
\textcolor{mBlue}{(Figure~\ref{fig:5-phases})} similar to that used in Ref.~\cite{basakSpatiallyDistributedRamp2023}. This protocol consists of two full hysteresis loops, called major loops, followed by six subloops and then two additional major loops. 
Figure~~\ref{fig:loops}(a) shows the result of our previous model~\cite{basakSpatiallyDistributedRamp2023} using above protocol, where the width of hysteresis loops is nonphysically small. 
Utilizing the \textcolor{mBlue}{C-}RFIM as described in Eq.~\eqref{eqn:Hamiltonian} without incorporating defect diffusion, we obtained the hysteresis loop shown in Figure~\ref{fig:loops}(b), which reproduces the hysteresis behavior of VO\(_2\) during a full temperature loop.

The outer loops in Figure~\ref{fig:loops} represent the major loop of the temperature protocol in Figure~\ref{fig:5-phases}. In this loop, the sample is heated from a fully insulating state to a fully metallic state and then cooled back to the fully insulating state. The inner loops correspond to the subloops, where the sample starts from the insulating state at a temperature \(T_\textnormal{L}\) , is heated up to a temperature \(T_\textnormal{H}\) within the phase coexistence region, and then cooled back to \(T_\textnormal{L}\).

\subsection{Ramp Reversal Memory}

When defect motion is introduced into the C-RFIM, ramp reversal memory emerges during the subloops \cite{basakSpatiallyDistributedRamp2023}. As depicted in Figure~\ref{fig:loops}(c), the subloops exhibit a counter-clockwise rotation with a decaying increment due to the dynamics of defects. After six subloops, the subsequent major loop (ML3) shows a significant increase in the insulating area at temperatures 
\textcolor{mBlue}{near}
\(T_\textnormal{H}\). If additional major loops are executed after these subloops, the RRM effect is gradually erased, and the hysteresis loop eventually returns to the shape observed in the initial major loop (ML1).

\textcolor{mRed}{The physical mechanism underlying RRM in our model is that when the sample is in the mixed phase having both metal and insulator domains, the thermodynamic tendency is for the metallic regions to have a higher defect concentration than the insulating regions.  This is similar to the floating melting zone refinement method for removing impurities from a crystal (for detailed microscopic explanations, see the Supplementary Information of \cite{basakSpatiallyDistributedRamp2023}).
This means that during the subloops, as metal repeatedly expands and recedes, it carries with it a higher concentration of defects each time.  The main driving force is different equilibrium defect densities in metallic versus insulating regions.
This redistribution locally modifies the critical temperature according to Eq.~\eqref{eqn:diffusion-segregation}, creating a ``memory'' of the thermal cycling history. With repeated minor loops, defects accumulate in certain regions progressively, enhancing the memory effect. However, when the sample undergoes a full major loop that becomes completely metallic, the defects can redistribute in a more uniform way throughout the sample, gradually resetting the accumulated memory and restoring the original hysteresis behavior.
The interactions introduced in Eqn.~\ref{eqn:Hamiltonian} have an indirect effect on the defect density through the morphology of the domains, set by a combination of interactions and local random field effects.
}


\subsection{\textcolor{mBlue}{Parameter} Dependence of the RRM Effect}

\begin{figure}[htb]
    \def\yposition{0.7}
    \centering
    \begin{subfigure}[b]{0.48\textwidth}
        \begin{tikzpicture}
            \node[anchor=south west, inner sep=0] (image) at (0,0) 
                {\includegraphics[width=\textwidth]{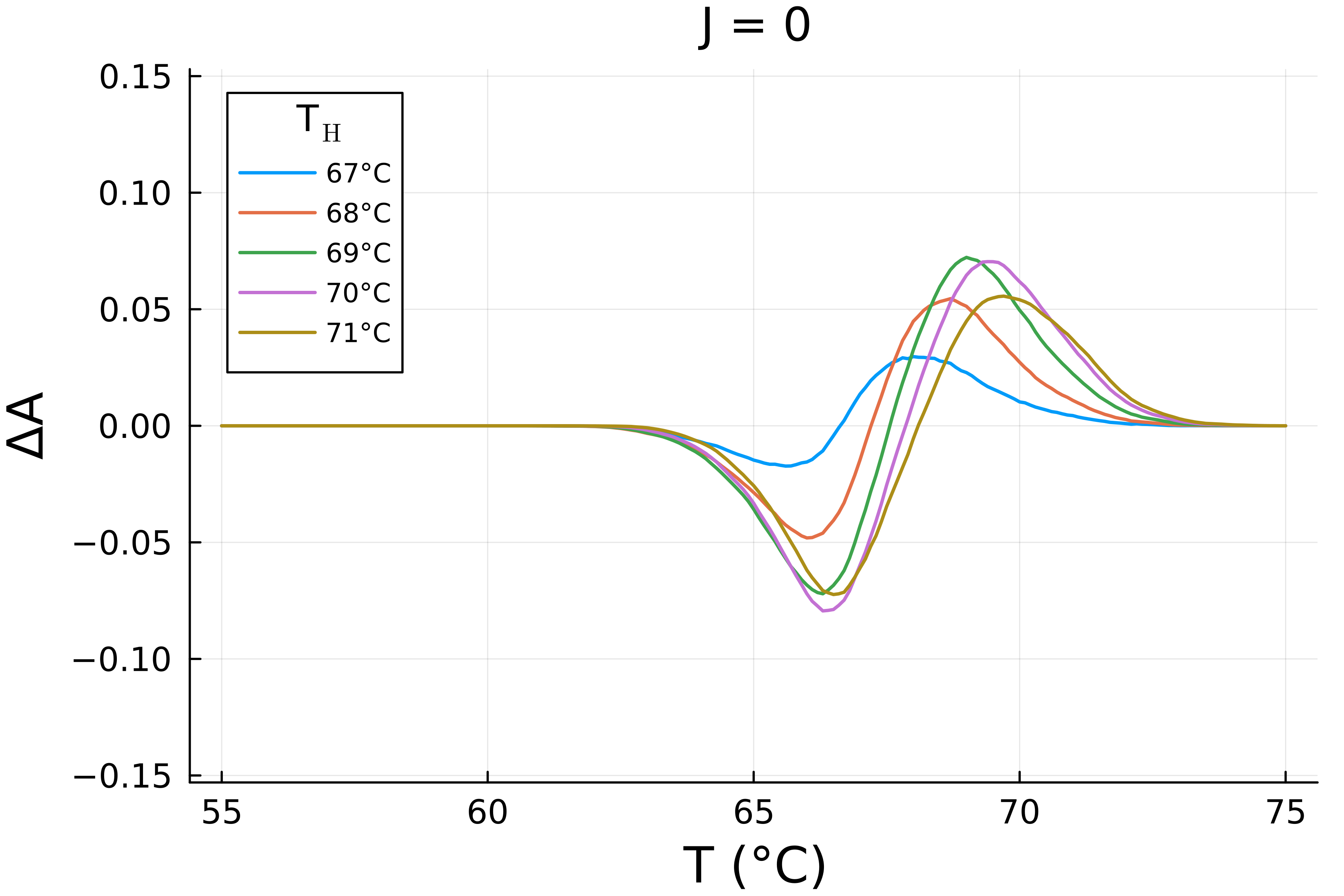}};
            \node[font=\bfseries, anchor=north west] at (0.05\textwidth,\yposition\textwidth) {(a)};
        \end{tikzpicture}
        \label{fig:R-diff_a}
    \end{subfigure}
    \hfill
    \begin{subfigure}[b]{0.48\textwidth}
        \begin{tikzpicture}
            \node[anchor=south west, inner sep=0] (image) at (0,0) 
                {\includegraphics[width=\textwidth]{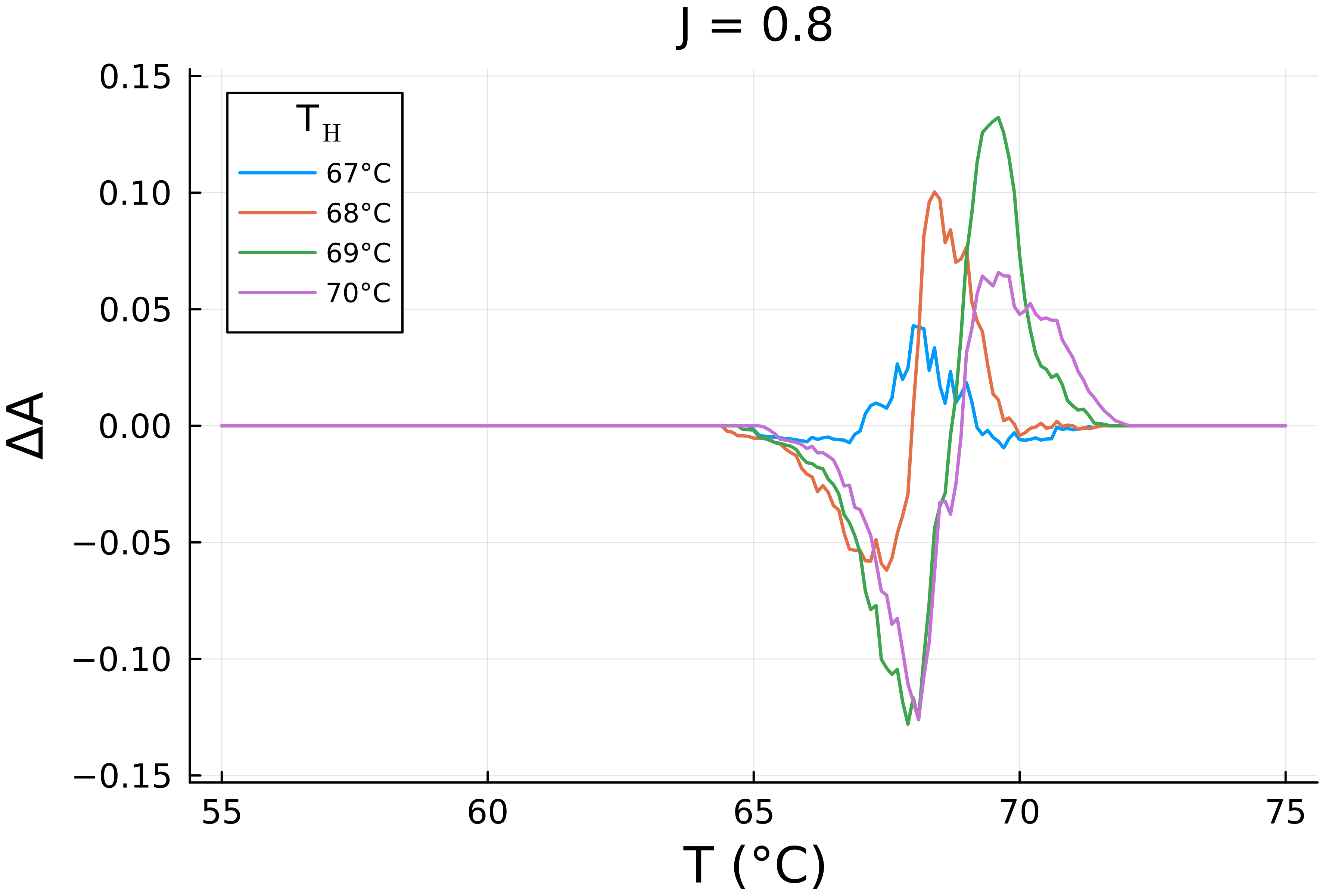}};
            \node[font=\bfseries, anchor=north west] at (0.05\textwidth,\yposition\textwidth) {(b)};
        \end{tikzpicture}
        \label{fig:R-diff_b}
    \end{subfigure}
    \caption{
    Simulated curves of fractional area change 
    versus temperature during the warming process. The difference of two heating curves \(\Delta A(T) = A_{\text{ML3}}(T) - A_{\text{ML2}}(T)\) is plotted, where \( A \) is the fraction of insulating area from Figure~\ref{fig:loops}. There are six temperature reversal subloops between Major Loop 2 and Major Loop 3, as shown in Figure~\ref{fig:5-phases}.
    (a) Without interactions (\( J = 0 \)), resulting in smoother curves.
    (b) With interactions (\( J = 0.8 \)), showing jaggedness due to avalanches, as well as enhanced memory due to interactions.
    }
    \label{fig:R-diff}
\end{figure}

Our model indicates that the magnitude of the maximum RRM effect depends on the highest temperature \( T_\textnormal{H} \) reached during the subloops
\textcolor{mBlue}{such that \( T_\textnormal{H} \) can be optimized to maximize the RRM effect.  
Figure~\ref{fig:R-diff} shows the change in relative insulating area before and after the subloops, $\Delta A(T) = A_{\rm ML3}(T) - A_{\rm ML2}(T)$.  
A peak in the change in relative insulating area before and
after the subloops 
happens with or without interactions.  
}
In the non-interacting case (Figure~\ref{fig:R-diff}(a)), the curves appear smooth, since Ising interactions that lead to avalanches are absent. Introducing interactions (\( J \neq 0 \)) enhances the RRM effect, and also introduces jaggedness in the curves due to these avalanches, as shown in Figure~\ref{fig:R-diff}(b).

\textcolor{mBlue}{To quantify the RRM effect, we define key parameters from the temperature-dependent area change curves in Figure~\ref{fig:R-diff}. For each curve, we identify the peak position (marked with a red dot in Figure~\ref{fig:annotate}) to define \( T^\textnormal{peak} \) as the temperature at which the peak occurs and \(\Delta A^\textnormal{peak} = \max(\Delta A(T))\) as the maximum change in insulating area.}

\subsubsection{\textcolor{mBlue}{Magnitude of Maximum RRM Effect}}

\begin{figure}[h]
    \centering
    \includegraphics[width=0.48\textwidth]{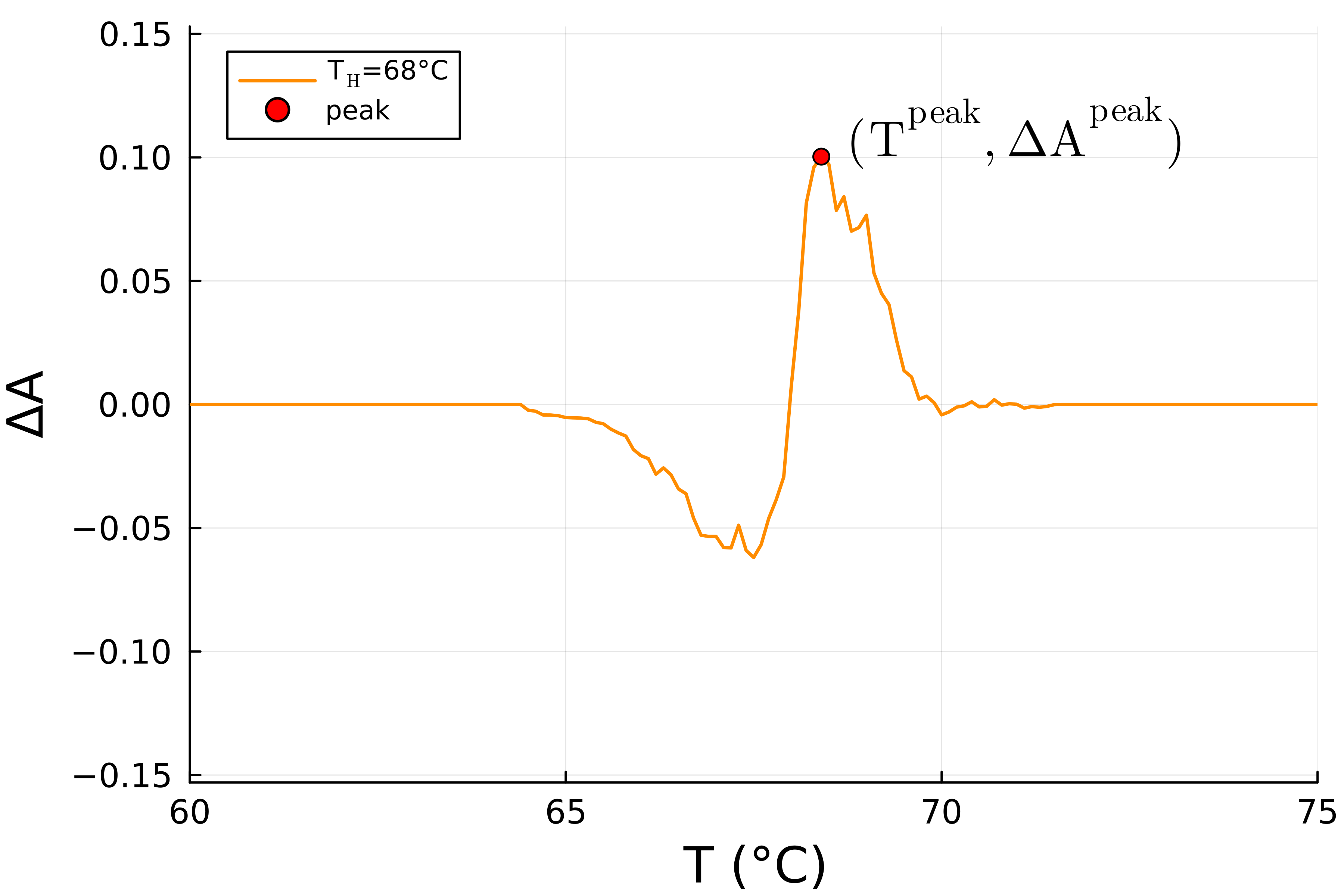}
    \caption{Definition of \(T^{\text{peak}}\) and \(\Delta A^{\text{peak}}\). The red point shows corresponding peak position for \( T_\textnormal{H} = \qty{68}{\degreeCelsius} \)}
    \label{fig:annotate}
\end{figure}

\begin{figure}[ht]
    \def\yposition{0.78}
    \centering
    \begin{subfigure}[b]{0.48\textwidth}
        \begin{tikzpicture}
            \node[anchor=south west, inner sep=0] (image) at (0,0) 
                {\includegraphics[width=\textwidth]{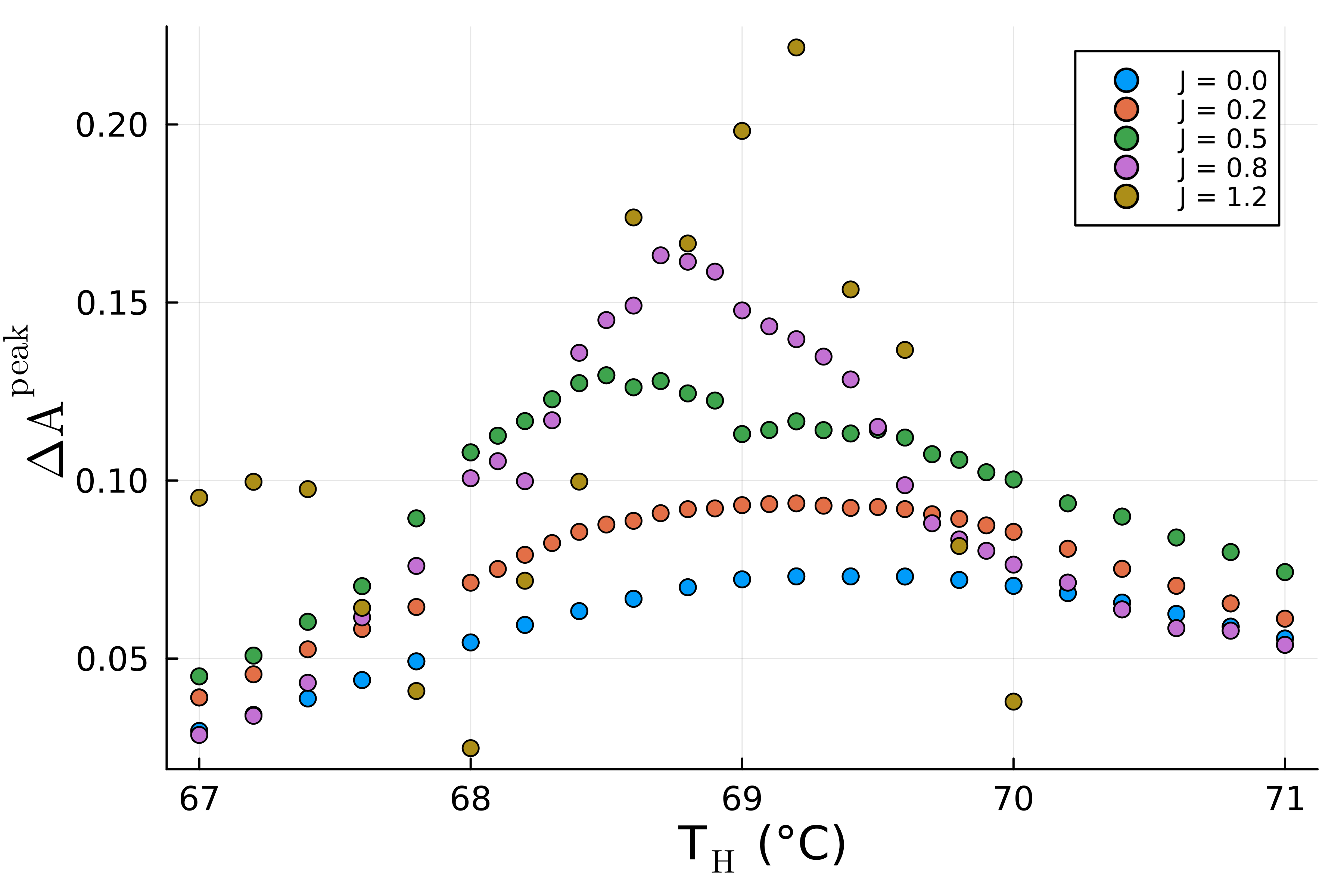}};
            \node[font=\bfseries, anchor=north west] at (0.05\textwidth,\yposition\textwidth) {(a)};
        \end{tikzpicture}
        \label{fig:Max_A_a}
    \end{subfigure}
    \hfill
    \begin{subfigure}[b]{0.48\textwidth}
        \begin{tikzpicture}
            \node[anchor=south west, inner sep=0] (image) at (0,0) 
                {\includegraphics[width=\textwidth]{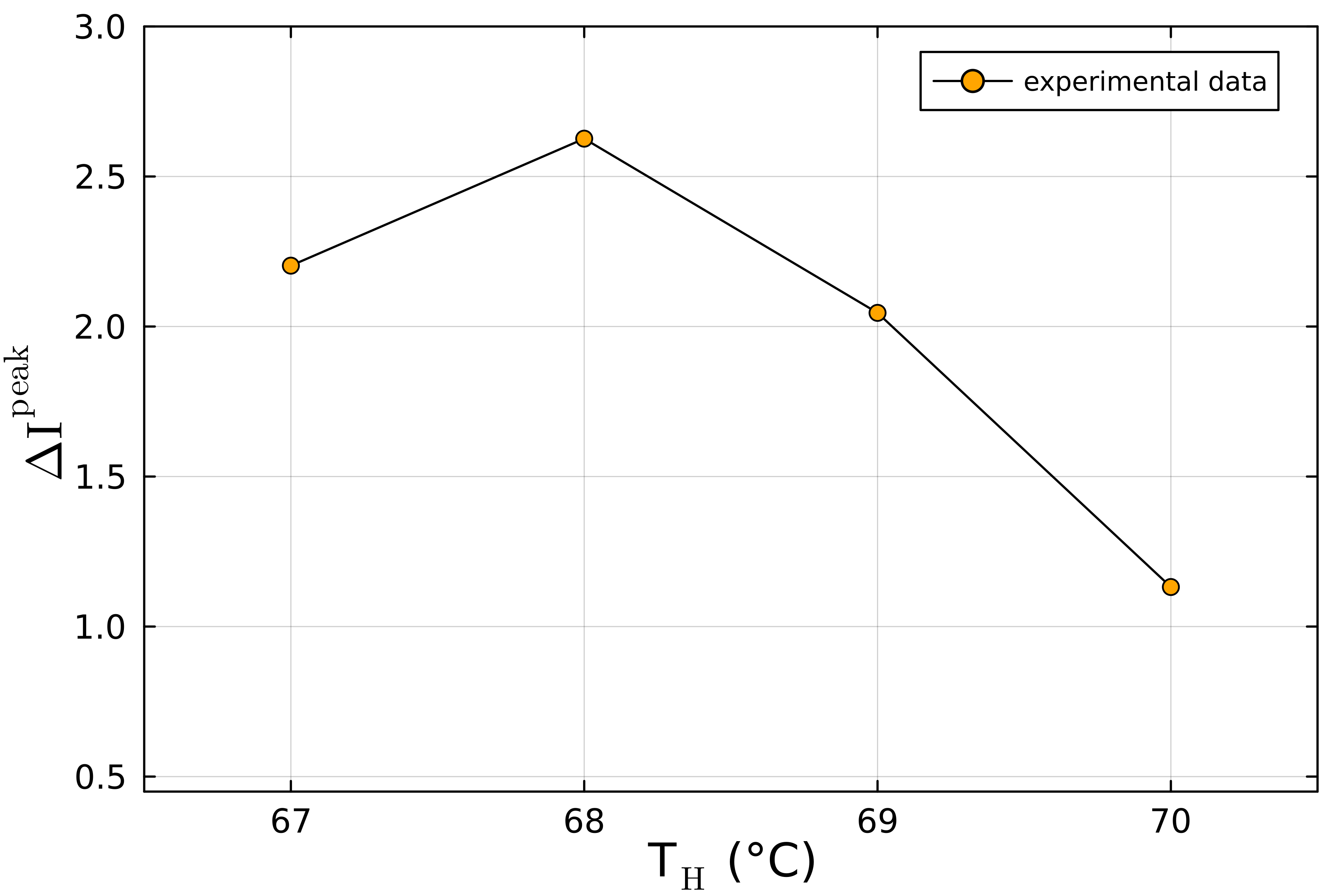}};
            \node[font=\bfseries, anchor=north west] at (0.05\textwidth,\yposition\textwidth) {(b)};
        \end{tikzpicture}
        \label{fig:Max_A_b}
    \end{subfigure}
    \caption{
        (a) Simulated RRM effect versus \( T_\textnormal{H} \);
        (b) Experimental optical data of RRM effect versus \( T_\textnormal{H} \). 
    Note that in both theory and experiment, we find a peak in the magnitude of the RRM effect as a function of T$_H$. 
    }   
    \label{fig:Max_A}
\end{figure}

\textcolor{mBlue}{We perform systematic simulations across different values of \(J\) and \(T_\textnormal{H}\) using our temperature protocol. For each parameter set, we calculate the difference between major loops immediately before and after the subloops as shown in Figure~\ref{fig:R-diff}, extract \(\Delta A^\textnormal{peak}\) from the resulting curves as defined in Figure~\ref{fig:annotate}, and plot the data using \((J, T_\textnormal{H}, \Delta A^\textnormal{peak})\) triplets.
The results are shown in Figure~\ref{fig:Max_A}(a). 
Our results reveal that not only is there a peak in the RRM effect as a function of \(T_\textnormal{H}\), but that
{\em interactions amplify the RRM effect}.  
}
\textcolor{mRed}{
Specifically, as interaction strength is increased the RRM effect grows in intensity, and the peak of the RRM effect as a function of \(T_\textnormal{H}\) becomes more pronounced.
This result reveals that interactions fundamentally change how memory accumulates. Without interactions, metal-insulator phase boundaries move gradually and continuously, 
creating defect depletion regions just outside of each metallic domain. 
The depletion region both raises the local \(T_c\) (making it harder for metal to advance there in subsequent warming ramps), and slows the flow of defects across the boundary, limiting the RRM effect.  However, we hypothesize that because interactions enable avalanches, boundaries can jump over depletion regions, allowing higher defect current across phase boundaries and increasing the memory accumulation during subloops. 
}

\textcolor{mBlue}{
We note that 
we observe the emergence of a possible double-peak structure in the relationship between the maximum RRM effect and \(T_\textnormal{H}\) (Figure~\ref{fig:Max_A}(a)). However, further investigation is needed to confirm the robustness of this double-peak feature, as it might be sensitive to the specific realization of the correlated random field
and associated noise.  
}

\textcolor{mBlue}{
In Figure~\ref{fig:Max_A}(b), 
we plot the maximum RRM effect vs. \(T_\textnormal{H}\)
derived from experiment, measured via
reflected light using a CCD camera.
We quantify the RRM effect in terms of the maximum difference
in the average intensity \(I\) of reflected light,
comparing  
major loops measured immediately before and immediately
after the temperature subloops:
\(\Delta I^{\rm peak}=\max(\Delta I)\), where \(\Delta I(T) = I_{\rm ML3}(T) - I_{\rm ML2}(T)\).
Additional details from the experiment are provided in the Supplementary Information, including color-coded VO\(_2\) sample maps showing the spatial distribution of RRM encoding.
Note the agreement between theory and experiment:  In both cases, we find that there is a peak in the magnitude of the RRM effect as a function of the subloop turnaround temperature \(T_\textnormal{H}\).  
}


\subsubsection{\textcolor{mBlue}{Optimal Temperature 
\( T^\textnormal{peak} \) for Maximum RRM Effect}}

\begin{figure}[htb]
    \centering
    \includegraphics[width=0.5\textwidth]{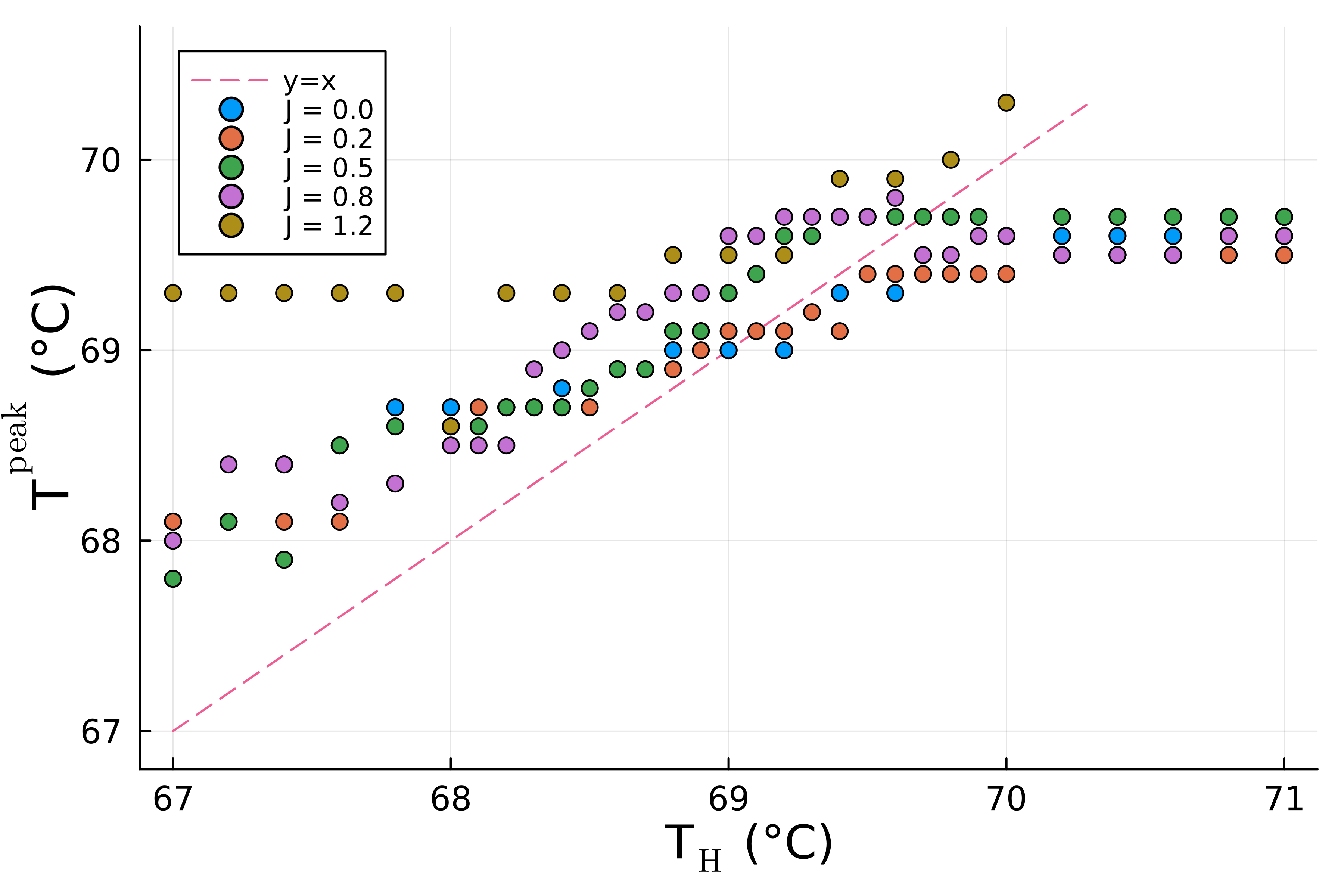}
    \caption{
        Simulated optimal temperature \textcolor{mRed}{
        \( T^\textnormal{peak} \) 
        for each \( T_\textnormal{H} \), as illustrated in Figure~\ref{fig:annotate}.}
        The dotted purple line is a guide to the eye.
    }
    \label{fig:opt_T}
\end{figure}

The optimal temperature for achieving the maximum RRM effect, denoted as 
\( T^\textnormal{peak} \),
also exhibits a clear dependence on 
\( T_\textnormal{H} \) (Figure~\ref{fig:opt_T}). 
Our simulations reveal that 
\( T^\textnormal{peak} \) 
closely tracks \( T_\textnormal{H} \) 
\textcolor{mRed}{for \( T_\textnormal{H} < 70\unit{\degreeCelsius} \) and saturates as  
\( T_\textnormal{H} \) approaches the temperature at which the hysteresis loop closes. 
This observed temperature dependence suggests that the defect distribution established by 
repeated movement of
domain walls at \( T_\textnormal{H} \) plays a crucial role in facilitating the memory effect. }
It corroborates findings from previous studies~\cite{anouchiUniversalityMicrostrainOrigin2022} and aligns with our experimental observations.

\subsubsection{\textcolor{mBlue}{Optimal Subloop Turnaround Temperature \(T_\textnormal{H}\)}}

\begin{figure}[htb]
    \centering
    \includegraphics[width=0.7\linewidth]{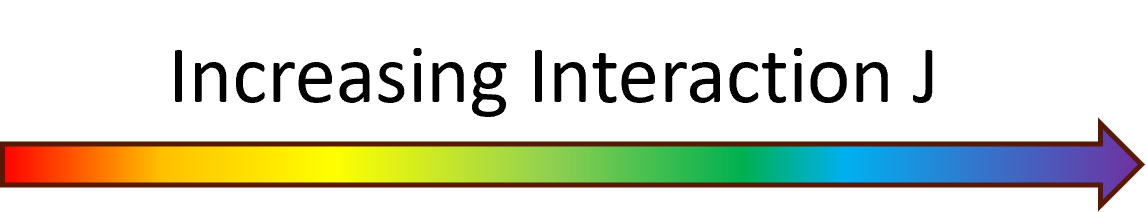}
    \includegraphics[width=\linewidth]{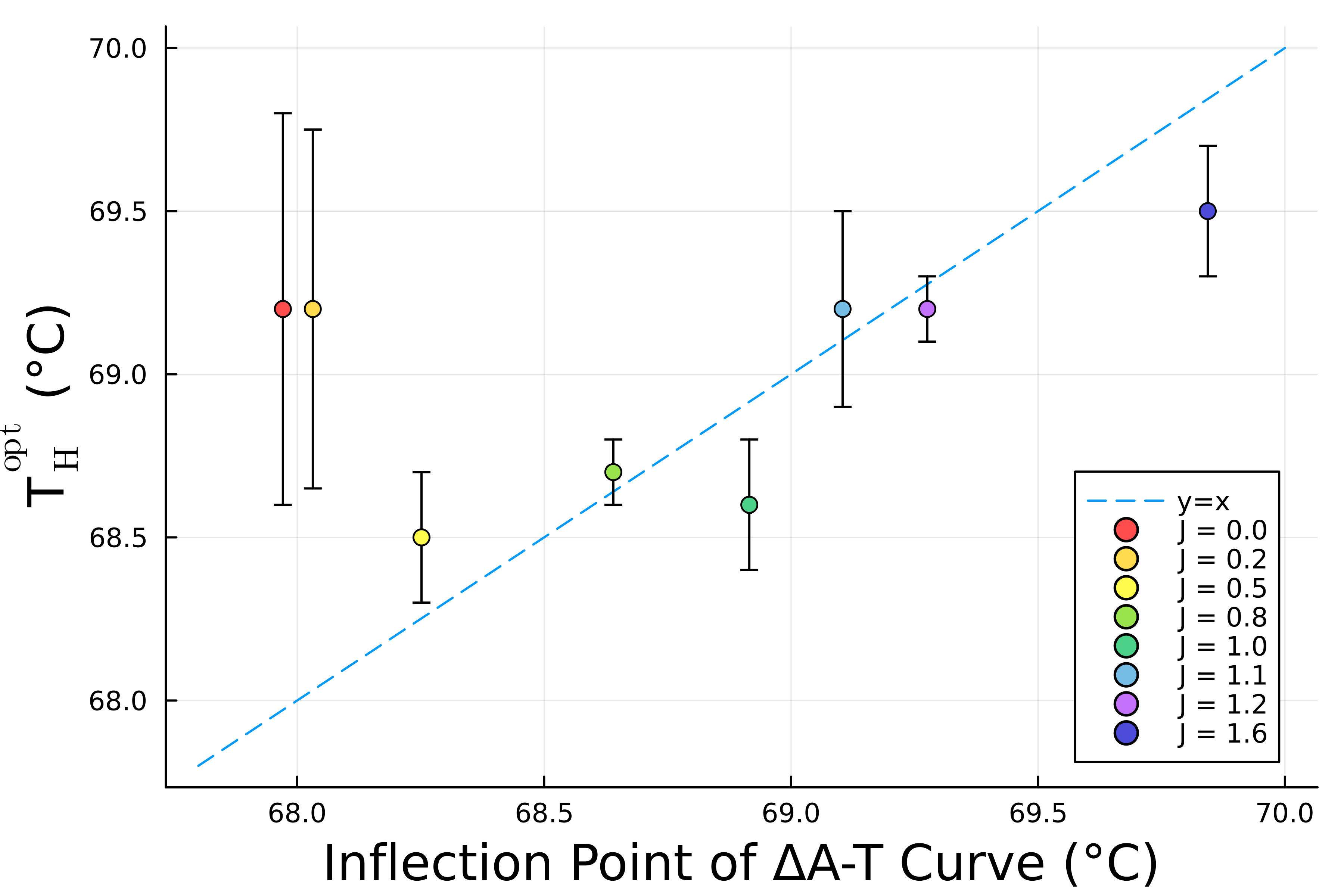}
    \caption{Optimal \(T_\textnormal{H}\) versus the inflection point of the warming branch of Major Loop 2, extracted from the simulation data. 
    \textcolor{mBlue}{
    Error bars indicate the precision of finding \(T_\textnormal{H}^\textnormal{opt}\) from Figure~\ref{fig:Max_A}(a).
    They are derived from the peak widths in Figure~\ref{fig:Max_A}(a), and set by the range of \(T_\textnormal{H}\) values where \(\Delta A^\textnormal{peak}(T_\textnormal{H}) > 0.95 \Delta A^\textnormal{peak}(T_\textnormal{H}^\textnormal{opt})\). }
    \textcolor{mRed}{
    For moderate to strong interactions, the inflection point increases with increasing \(J\).  The dotted blue line is a guide to the eye.}   
    }
    \label{fig:inflection}
\end{figure}

Since \(T_\textnormal{H}\) is a readily controllable parameter in experiments, understanding its influence on the RRM effect is paramount for optimizing device performance.  Our results show that as the interaction strength \(J\) increases, the optimal \( T_\textnormal{H} \) for maximizing the RRM effect approaches the inflection point of the warming branch of the initial major hysteresis loop (ML1-W), as illustrated in Figure~\ref{fig:inflection}. This provides a valuable guideline for experimentalists seeking to maximize the RRM effect: the subloops should be designed such that \(T_\textnormal{H}\) is near the inflection point of the major loop's warming branch.

\begin{figure*}[htb]
    \centering
    \begin{tikzpicture}
        \node[anchor=south west, inner sep=0] (image) at (0,0) 
            {\includegraphics[width=0.99\linewidth]{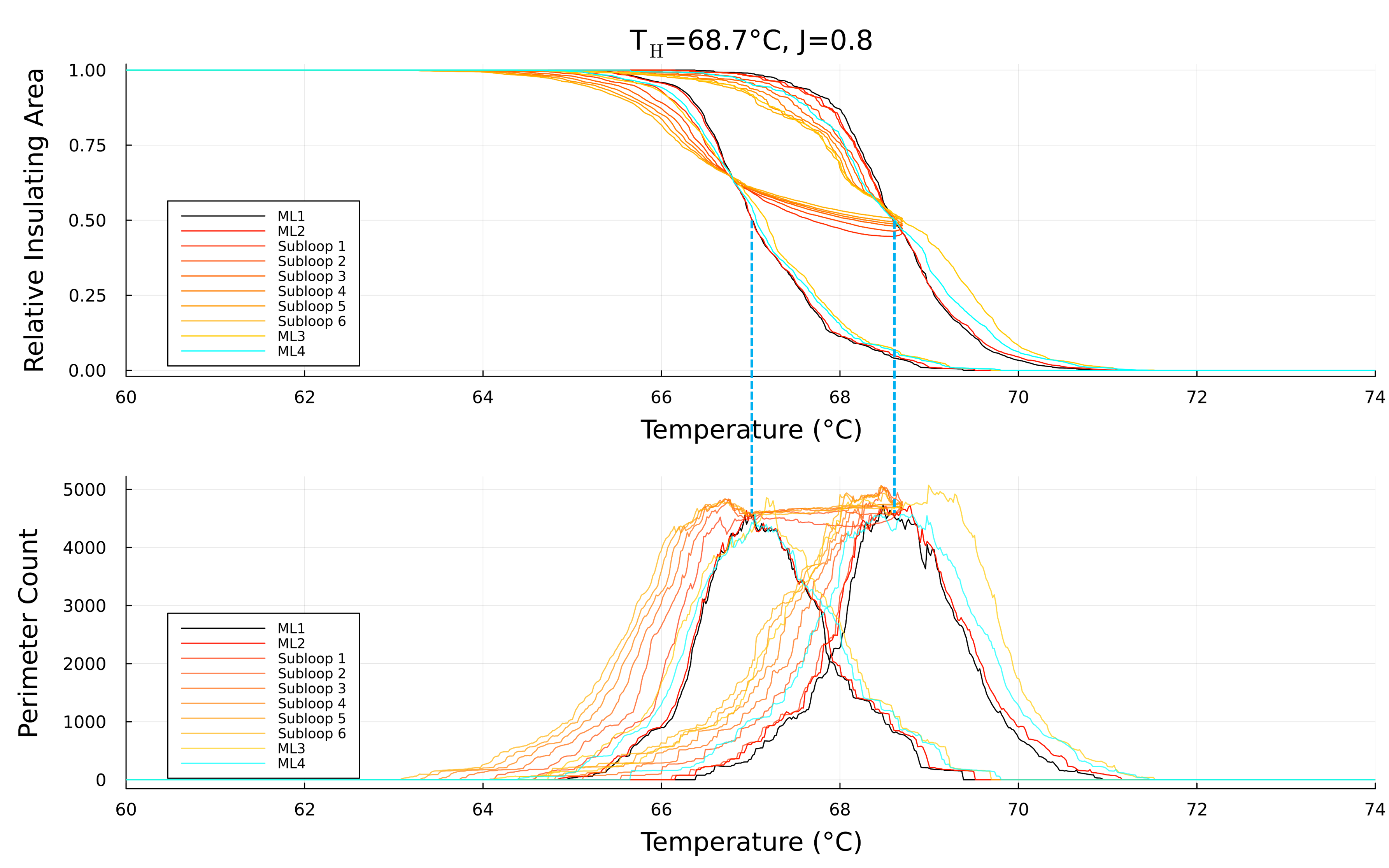}};
    \end{tikzpicture}
    \caption{Comparison of temperature-dependent phase transition behaviors, showing the relationship between the metallic/insulating area fraction and the total metal-insulator boundary length (perimeter) at \textcolor{mBlue}{\(J=0.8\), \(T_\textnormal{H}=T_\textnormal{H}^\textnormal{opt}(J\!=\!0.8)=68.7\) (purple dot in Figure~\ref{fig:inflection})}. Vertical dashed lines indicate the inflection points of the warming (right line) and cooling (left line) branches of the hysteresis loop, corresponding to the temperatures at which the total metal-insulator boundary length is maximized. In the absence of interactions, the perimeter curve would be \textcolor{mBlue}{approximately} proportional to the absolute value of the derivative of the relative area curve.}
    \label{fig:vertical_compare}
\end{figure*}

\textcolor{mRed}{The roughly linear relationship observed in Figure~\ref{fig:inflection} for moderate to strong interactions
can be understood by considering the total perimeter between metallic and insulating domains.}
As shown in Figure~\ref{fig:vertical_compare}, the system exhibits a maximum in the total metal-insulator boundary length near the inflection point of the warming branch. 
\textcolor{mBlue}{
Since the defect movement responsible for the RRM effect increases with boundary length, maximizing the duration of the subloops near the inflection point promotes enhanced memory accumulation. 
}
At temperatures significantly below the inflection point, memory accumulation is slow due to the limited interfacial area. Conversely, at temperatures significantly above the inflection point, the established memory effect is diminished by defect diffusion and the reduction in the minority phase. The combined effect results in an optimal turn-around temperature, \(T_\textnormal{H}^\textnormal{opt}\), that sits near the inflection point.

It is crucial to distinguish between different definitions of the ``inflection point.'' The relevant inflection point for maximizing RRM, as described above, is that of the insulator-metal fraction versus temperature curve. This can be determined by tracking the grayscale in optical data~\cite{AlzateBanguero2025, basakSpatiallyDistributedRamp2023}. \textcolor{mBlue}{For data used in Figure~\ref{fig:inflection}, we determine these inflection points by fitting the relative insulating curve with a hyperbolic tangent function.}  
However, the inflection point commonly reported in the literature is deduced from resistivity. This one can be misleading, as it is shifted with respect to the insulator-metal fraction due to complex percolation paths in the sample.

\newpage 

\subsection{\textcolor{mBlue}{What our model does not capture}}
\noindent

\textcolor{mRed}{
When comparing our theoretical results to experimental observations, we find that while our simulations capture several key features observed experimentally, certain discrepancies are present. Specifically, in Figure~\ref{fig:loops}(c), our simulations show a decrease in the insulating area at low temperatures, which is not seen in experimental hysteresis loop measurements.} This discrepancy may arise from 
our assumption  
that the total number of defects is conserved and that changes in \(T_{c}\) depend linearly on variations in the defect density \(\rho\). Our assumption leads to 
a sum rule
\(\left(\sum_i \Delta T_{c}^{(i)}=-\alpha\sum_i \Delta \rho_{i}=0\right) \). 
Consequently, if a portion of the hysteresis curve shifts upward, another portion must shift downward, which differs from experimental observations where the curves primarily shift upward.

In fact, density functional theory calculations indicate that the defects in VO\(_2\) are most likely oxygen vacancies and interstitials~\cite{cuiFormationEnergiesIntrinsic2016}, whose concentrations depend on temperature. 
\textcolor{mRed}{Since the experiments are carried out with the samples exposed to the air,
it seems reasonable that oxygen vacancies and interstitials should change with temperature.
At lower temperatures, a reduction in defect density could, for example, counteract the downward shift of the hysteresis curve in the low-temperature regime. 
We leave the exploration of this effect to future work.}
Another possibility involves non-point defects, such as phase boundaries and strain fields, which may increase the local critical temperature. 

It should also be noted that while our model of defect diffusion within dynamic interacting phase separated domains includes certain phenomenology of the scar model (that significant local action happens at phase boundaries), our model also includes phenomenology
that goes beyond the expectations of the scar model, including:
(1) Memory accumulates throughout the entire bulk of the sample, and
(2) Whereas some regions will have increased local transition temperature, some other regions will have decreased local transition temperature. Future work may also consider integrating our defect motion model with the scar model proposed by \textcite{vardiRampReversalMemoryPhaseBoundary2017}, with the goal of developing a more comprehensive theoretical framework.

\subsection{Future Outlook and General Applicability}
Our model makes minimal assumptions:  that 
phase separated domains interact, and that for thermodynamic reasons point defects preferentially gather in one type of domain over the other.  
As a result, our theoretical model predicts that RRM should be a 
{\em general phenomenon} in materials undergoing patterned electronic phase separation, as conjectured in Ref.~\onlinecite{fried2025NewMemoryEffect}.
This is consistent with the fact that the RRM effect has been seen in a variety of
TMO's, including VO$_2$ as discussed extensively here,
as well as V$_2$O$_3$ and NdNiO$_3$.\cite{vardiRampReversalMemoryPhaseBoundary2017,anouchiUniversalityMicrostrainOrigin2022}
The minimal assumptions of our theoretical model imply that the RRM phenomenon should be far more widespread, and should be observable in materials
with locally phase separated electronic domains (as recently found in  1T-TaS$_2$~\cite{fried2025NewMemoryEffect}).
Potential candidates include:
cuprate high temperature superconductors, some of which have
shown regimes of interleaved phase separated regions of superconducting and non-superconducting domains~\cite{Gomes_Visualizing_Pair_Formation_Bi2212_2007, tromp-puddles}; graphene~\cite{Martin_ElectronHole_Puddles_Graphene_2008}, strontium iridates~\cite{milan-iridates,madhavan-iridate}, 2DEG superconductors,~\cite{Bert_Magnetic_Imaging_LAO_STO_2011, Biscaras_Multiple_QC_2D_SC_2013} manganites,~\cite{moreo-dagotto-manganites} and even
iron-based superconductors~\cite{allan-pnictide}.




\section{Conclusion}

In order to address the experimental findings not fully explained by the scar model, we developed the non-interacting defect motion model in our previous work. This model successfully reproduced several key observations, including decreased \(T_c\) values and RRM effects that accumulate deep inside metal/insulator patches.  However, 
due to the lack of interactions among metal and insulator domains, it did not include avalanche physics and the hysteresis loops were not very wide.  
To improve our theoretical framework, we have coupled the defect motion model with a Correlated Random Field Ising Model (C-RFIM) on a site-by-site basis.
\textcolor{mRed}{Introducing Ising interactions provides several major advantages that transform our model from a qualitative explanation to a quantitative predictive tool. First, the interacting model successfully reproduces the complete experimental hysteresis width (Figure~\ref{fig:loops}(c)) and captures avalanche behavior evident in the jaggedness of Figure~\ref{fig:R-diff}(b), which the non-interacting model cannot reproduce. 
Most significantly, we discover that interaction strength directly controls memory performance: increasing the nearest-neighbor interaction \(J\) dramatically enhances the maximum RRM effect (Figure~\ref{fig:Max_A}(a)). 
This discovery provides a clear pathway for optimizing memory devices through material parameters that control effective interaction strength.}
Notably, our interacting model predicts that the maximum RRM effect occurs near the inflection point of the warming process, consistent with the optical measurements on VO$_2$ reported here. 

By incorporating both defect dynamics and interactions between metallic and insulating domains, our model provides valuable insights into the underlying mechanisms driving these phenomena. 
These findings not only advance our fundamental understanding of RRM but also establish a theoretical foundation for developing more effective memory devices, neuromorphic computing applications, and other potential technologies based on this material.
Moreover, the minimal assumptions of our theoretical model imply that the
RRM effect should be a generic phenomenon in materials exhibiting
local electronic phase separation. 

\begin{acknowledgments}
We acknowledge helpful conversations with K.~A.~Dahmen and A. Sharoni. E.W.C. acknowledges support from NSF Grant No.~DMR-2006192 and from a Fulbright Fellowship, and thanks the Laboratoire de Physique et d'\'{E}tude des Mat\'{e}riaux (LPEM) at \'{E}cole Sup\'{e}rieure de Physique et de Chimie Industrielles de la Ville de Paris (ESPCI) for hospitality.
E.W.C. and Y.S. acknowledge support from DOE BES Award No.~DE-SC0022277.
This research was supported in part through computational resources provided by Research Computing at Purdue, West Lafayette, Indiana~\cite{rcac-purdue}.
The work at UCSD (PS, IKS)  was supported through an Energy Frontier Research Center program funded by the US Department of Energy (DOE), Office of Science, Basic Energy Sciences, under Grant DE-SC0019273.
The work at ESPCI (M.A.B., L.A., and A.Z.) was supported by
Cofund AI4theSciences hosted by PSL University, through the European Union's Horizon 2020 Research and Innovation Programme under the Marie Skłodowska-Curie Grant No.~945304.
\end{acknowledgments}







\bibliography{VO2references,other-materials}

\end{document}



\title{\LARGE Supporting Information \\
\large Effects of Interactions and Defect Motion on Ramp Reversal Memory in Locally Phase Separated Materials}

\author{Y.~Sun} 
\affiliation{%
Department of Physics and Astronomy, \\Purdue University, West Lafayette, IN 47907, USA
}%
\affiliation{
Purdue Quantum Science and Engineering Institute, West Lafayette, IN 47907, USA
}

\author{M.~Alzate Banguero}
\affiliation{%
Laboratoire de Physique et d'\'Etude des Matériaux, ESPCI Paris, France 
}%
\affiliation{%
Université, CNRS, Sorbonne Université, 75005 Paris, France 
}

\author{P.~Salev}
\affiliation{%
Department of Physics and Astronomy, University of Denver, Denver, Colorado 80208, USA
}%

\author{Ivan~K.~Schuller}
\affiliation{%
Department of Physics and Center for Advanced Nanoscience, University of California-San Diego,
La Jolla, California 92093, USA
}%

\author{L.~Aigouy}
\affiliation{%
Laboratoire de Physique et d'\'Etude des Matériaux, ESPCI Paris, France 
}%
\affiliation{%
Université, CNRS, Sorbonne Université, 75005 Paris, France 
}

\author{A.~Zimmers}
\email{azimmers@espci.fr}
\affiliation{%
Laboratoire de Physique et d'\'Etude des Matériaux, ESPCI Paris, France 
}%
\affiliation{%
Université, CNRS, Sorbonne Université, 75005 Paris, France 
}

\author{E.~W.~Carlson}
\email{ewcarlson@purdue.edu}
\affiliation{%
Department of Physics and Astronomy, \\Purdue University, West Lafayette, IN 47907, USA
}%
\affiliation{
Purdue Quantum Science and Engineering Institute, West Lafayette, IN 47907, USA
}

\date{\today}


\maketitle



\section{Generating correlated random field}

To generate spatially correlated random fields, we employ Cholesky decomposition of the correlation matrix. For sites $y_i$ and $y_j$ on a 2D lattice, we assume exponential spatial correlation:
\[
\braket{y_i, y_j} = C_{ij} \sim \exp\!{\big(-\lvert\vec{r}_i-\vec{r}_j\rvert/\xi\big)}
\]

We express $y_i$ as a linear combination of independent standard Gaussian random variables $x_k$ (ignore summation symbol): $y_i = A_{ik} x_k$. The covariance matrix of vector $\vec{y}$ becomes:
\begin{align*}
\braket{y_i,\,y_j} &= \braket{A_{ik}x_k,\,A_{jl}x_l}\\
    &= A_{ik}A_{jl}\braket{x_k,\,x_l}\\
    &= A_{ik}A_{jl} \delta_{kl}\\
    &= A_{ik}A_{jk}\\
    &= A_{ik}(A^{T})_{kj}\\
    &= \big(AA^T\big)_{ij}
\end{align*}

\begin{figure}[htb]
    \centering
    \includegraphics[width=0.9\linewidth]{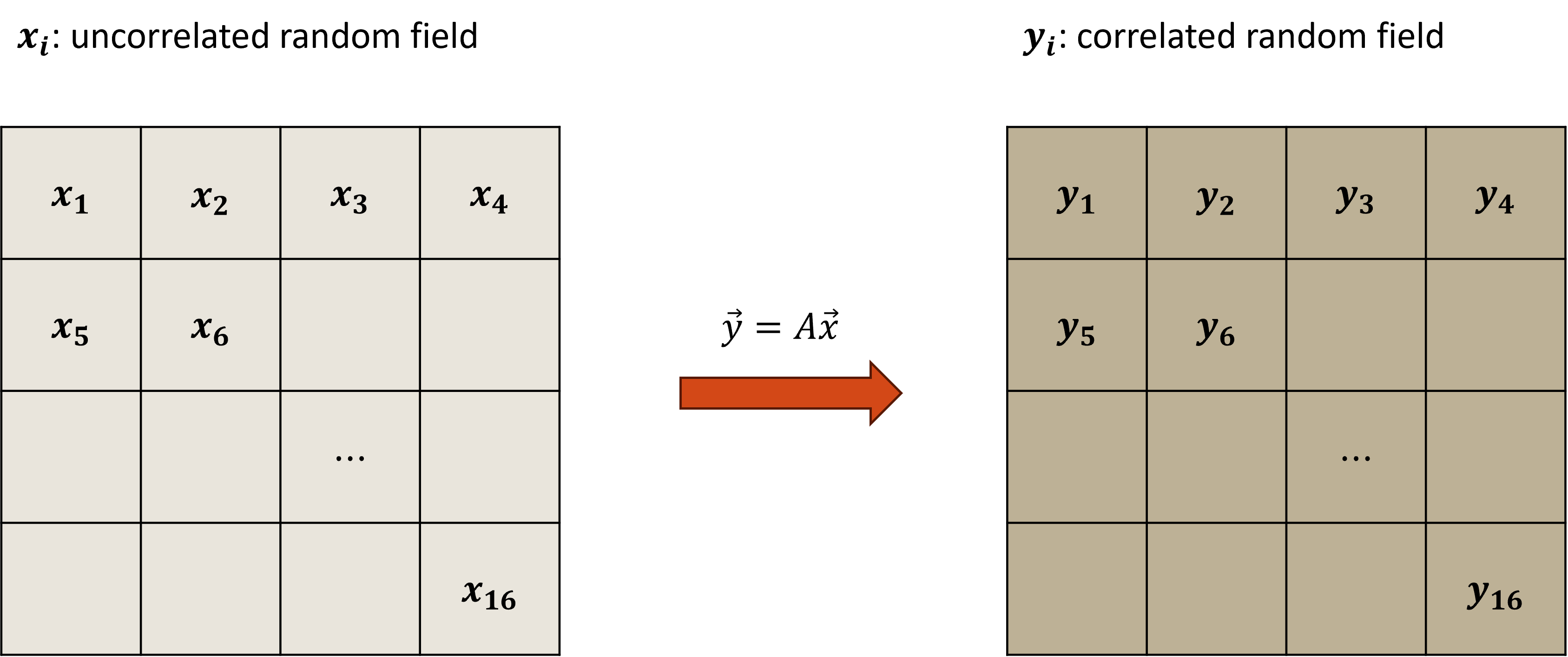}
    \caption{
    \textcolor{mRed}{Schematic configuration of \(y\) and \(x\).}
    }
    \label{fig:SI-y=Ax}
\end{figure}

The matrix $A$ satisfying $C = AA^T$ is obtained through Cholesky decomposition, yielding a lower triangular matrix $L$ such that $C = LL^T$. Setting $A = L$, we construct the correlated field via $\vec{y} = A\vec{x}$.

\begin{figure}[hbt]
    \centering
    \begin{picture}(0,0)
    \put(-10,140){{\footnotesize (a)}}
    \end{picture}
    \includegraphics[width=0.48\textwidth]{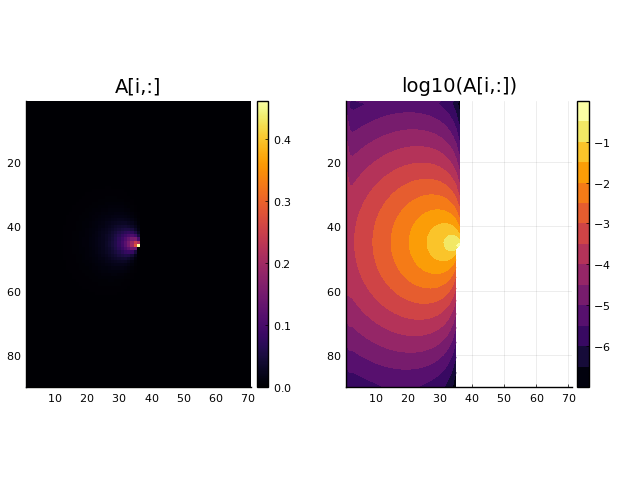}
    \begin{picture}(0,0)
    \put(-10,140){{\footnotesize (b)}}
    \end{picture}
    \includegraphics[width=0.48\textwidth]{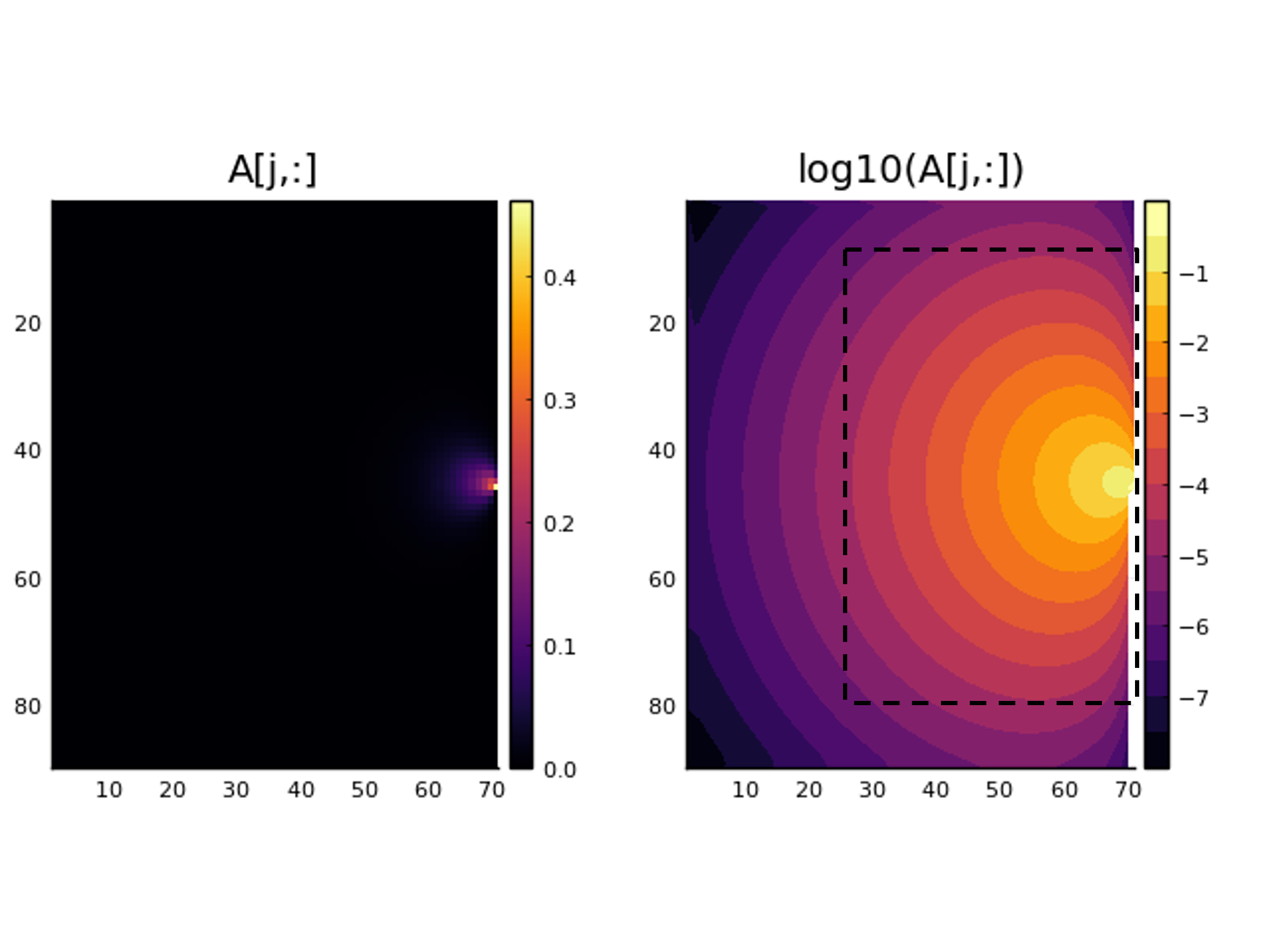}
    \caption{Visualization of selected rows from matrix \(A\). Rows are reshaped to 2D maps to show the spatial correlation. 
    (a) Theoretical correlation pattern between a center point \(y_i\) and all points in the uncorrelated field \(x\), shown in both linear scale (left) and logarithmic scale (right). The uncolored regions in the logarithmic plot correspond to elements that are exactly zero.
    (b) Theoretical correlation pattern between a middle-right edge point \(y_j\) and all points in the uncorrelated field \(x\), again in linear scale (left) and logarithmic scale (right), demonstrating the spatial variation in the correlation structure.
    The dashed bounding box is a schematic plot for the cutoff.}
    \label{fig:A_logA}
\end{figure}

Figure~\ref{fig:A_logA} demonstrates that correlation functions are approximately translationally invariant, enabling efficient implementation through convolution with a localized kernel. 
We apply a finite size cutoff to the \(j\)-th row, \(A[j,:]\), in Figure~\ref{fig:A_logA}(b) to obtain a localized 2D Cholesky kernel which ignores almost-zero elements far from the middle-right point. The kernel is then convolved with the 2D uncorrelated random field \(x_i\) to generate an approximation of the correlated random field \(y_i\). 
The main limitation of using a cutoff is that the finite kernel size restricts this approach to correlations within the cutoff distance, preventing generation of truly long-range correlated random fields.

Alternative matrix \(A'\) exists (e.g., using singular value decomposition) such that \(C=A'A'^T = AOO^TA^T\) where \(O\) is orthogonal, but these \(A'\) may require non-local combinations incompatible with spatial cutoffs.

\section{Gaussian Blurring Effect in Optical Measurement}

\begin{figure*}[bt]
    \centering
    \def\xposition{-10}
    \def\yposition{95}
    \def\mywidth{0.32}
    
    \begin{picture}(0,0)
    \put(\xposition,\yposition){{\footnotesize (a)}}
    \end{picture}
    \includegraphics[width=\mywidth\textwidth]{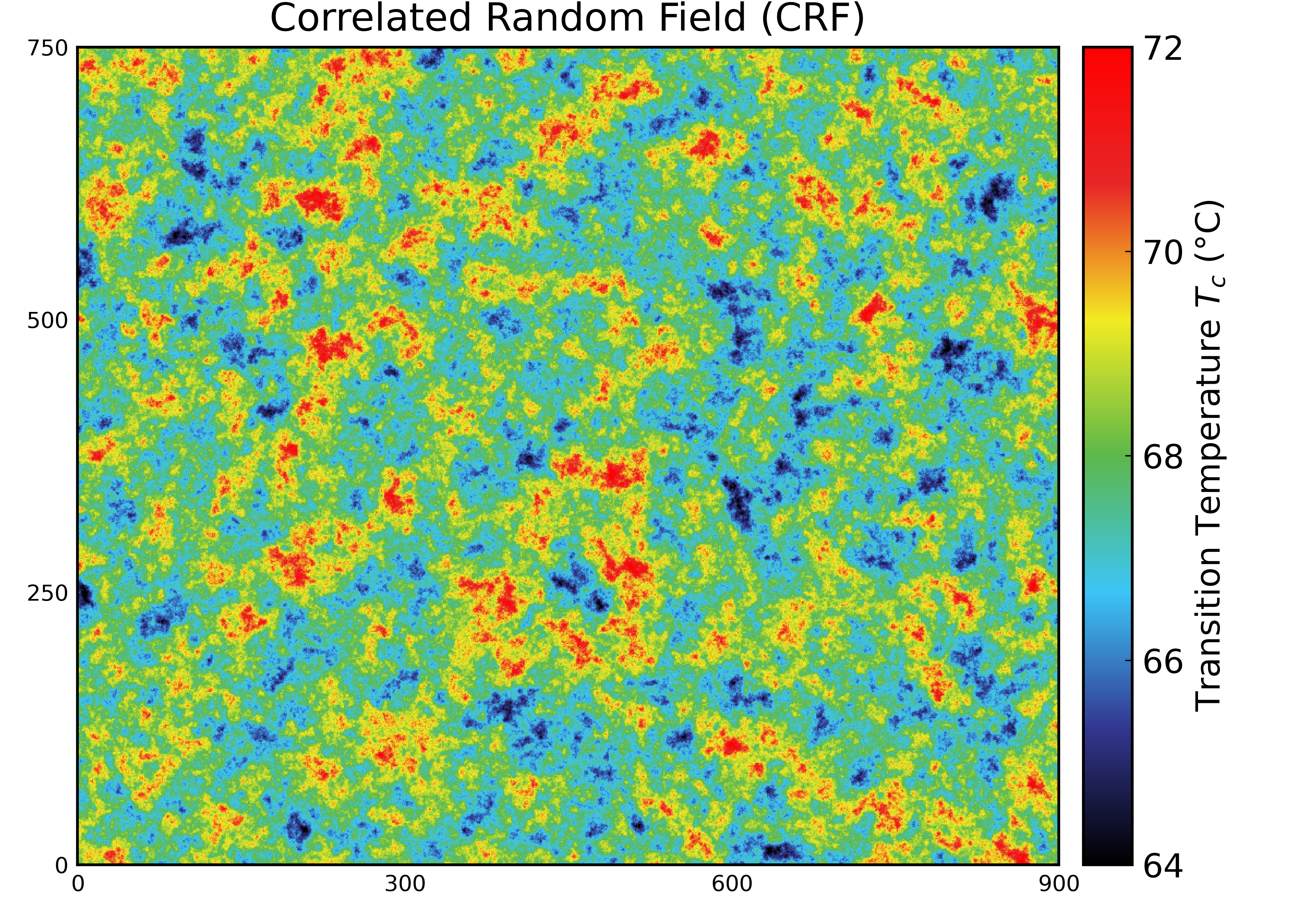}
    \begin{picture}(0,0)
        \put(\xposition,\yposition){{\footnotesize (b)}}
    \end{picture}
    \includegraphics[width=\mywidth\textwidth]{images/Tc_CRF_blurred.png}
    \begin{picture}(0,0)
        \put(\xposition,\yposition){{\footnotesize (c)}}
    \end{picture}
    \includegraphics[width=\mywidth\textwidth]{images/Tc_exp_TH70_run.png}\\
    \begin{picture}(0,0)
        \put(\xposition,\yposition){{\footnotesize (d)}}
    \end{picture}
    \includegraphics[width=\mywidth\textwidth]{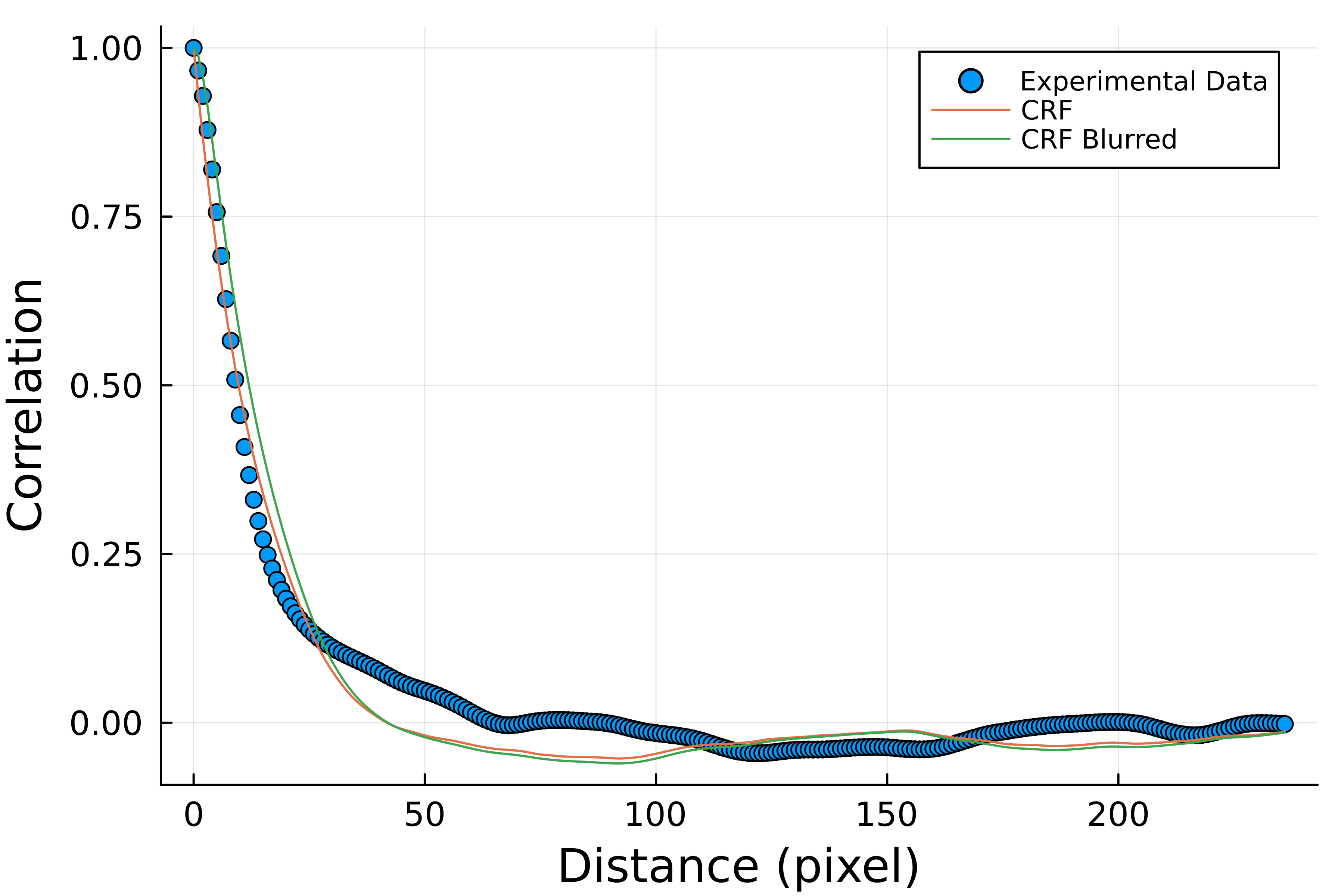}
    \begin{picture}(0,0)
        \put(\xposition,\yposition){{\footnotesize (e)}}
    \end{picture}
    \includegraphics[width=\mywidth\textwidth]{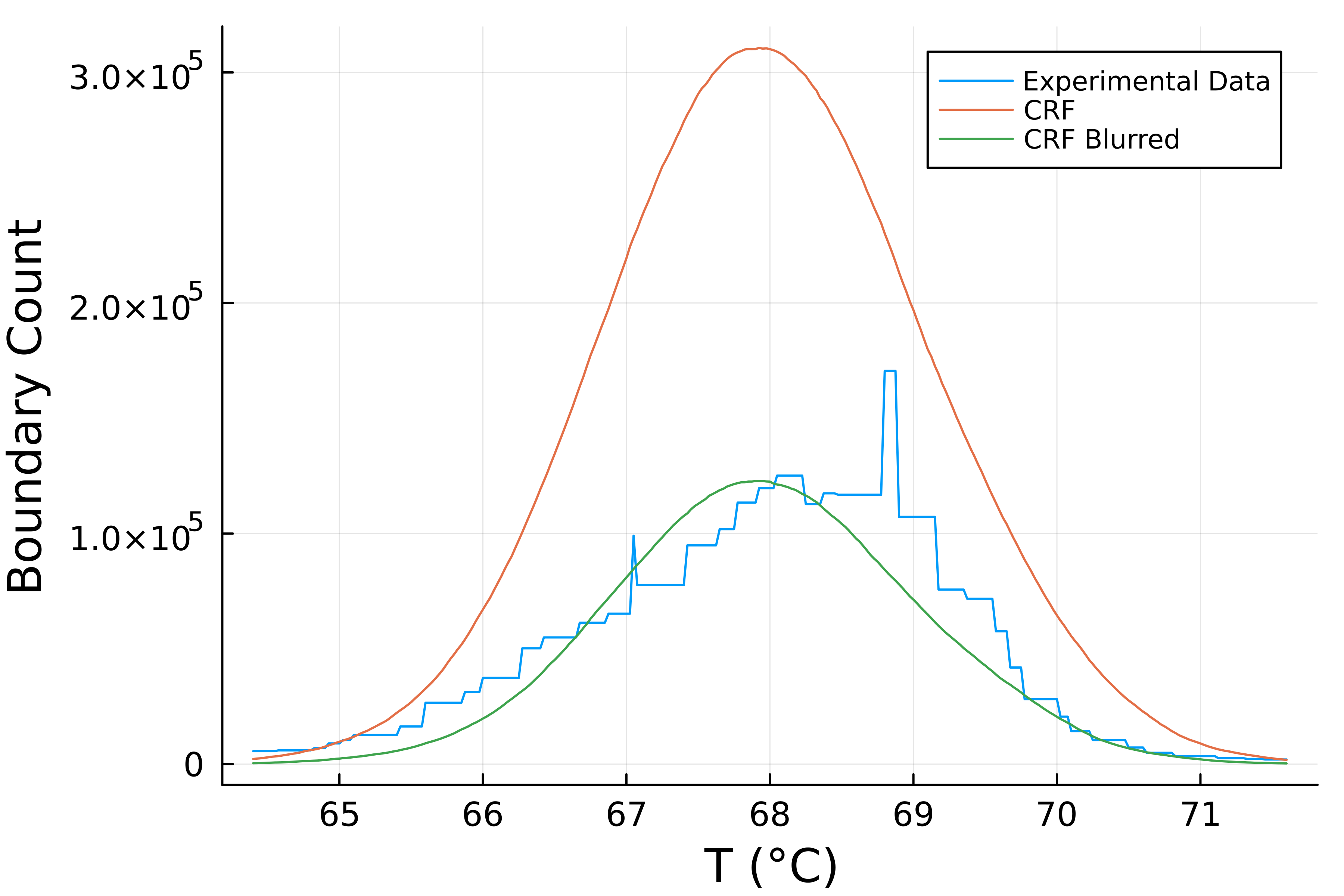}
    \begin{picture}(0,0)
        \put(\xposition,\yposition){{\footnotesize (f)}}
    \end{picture}
    \includegraphics[width=\mywidth\textwidth]{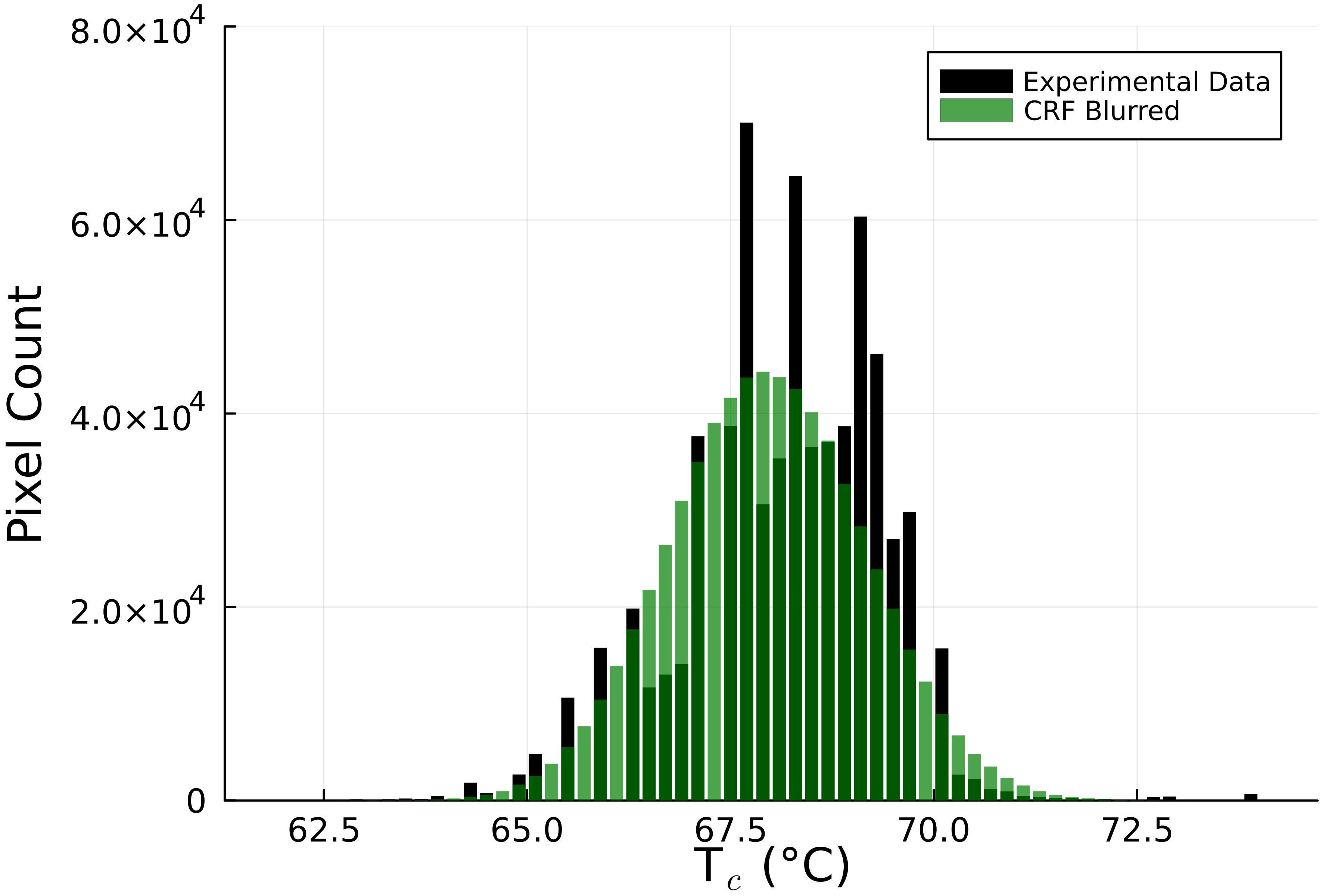}
    \caption{
        (a) Theoretical \(T_{c}\) map using correlated random field (CRF).
        (b) Theoretical \(T_{c}\) map using CRF with Gaussian blurring effect to facilitate comparison.
        (c) Experimental \(T_{c}\) map measured during the first warming process (ML1-W).
        (d) Correlation functions along one direction of experimental \(T_{c}\) map and different theoretical maps.
        (e) Phase boundary counted using initial \(T_c\) maps.
        (f) \(T_c\) distribution calculated using initial \(T_c\) maps.
    }
    \label{fig:SI-corRFIM}
\end{figure*}




As depicted in Figure~\ref{fig:SI-corRFIM}(e), the results obtained from optical measurements are approximately half those derived from the correlated random field. We hypothesize that this discrepancy arises from the Gaussian blurring effect, which effectively averages the values of neighboring pixels.  Detailed calculations reveal that Gaussian blurring introduces a parabolic rounding effect in the correlation function at small \(r\) (\(r \sim 0\)). This hypothesis is corroborated by the observation of a parabolic asymptotic behavior near \(r=0\) in Figure~\ref{fig:SI-corRFIM}(d). The curve representing the blurred correlated random field closely aligns with the experimental data at short distances. The non-vanishing tail observed in the experimental data can be attributed to the residual correlation characteristic of an ordered phase. Figure~\ref{fig:SI-corRFIM}(f) shows \(T_c\) distributions of experimental results and our Gaussian blurred correlated random field.

\section{Other Parameters in the Model}

The function \( f \) in C-RFIM is a monotonic increasing function satisfying \( f(0) = 0 \) and \( f'(0) = 1 \); under the normalization \( f'(0) = 1 \), scaling factors are absorbed into \( J \). The data presented in the main text uses \( f(x) = x \) for simplification.

\begin{figure}[h]
    \centering
    \includegraphics[width=0.98\textwidth]{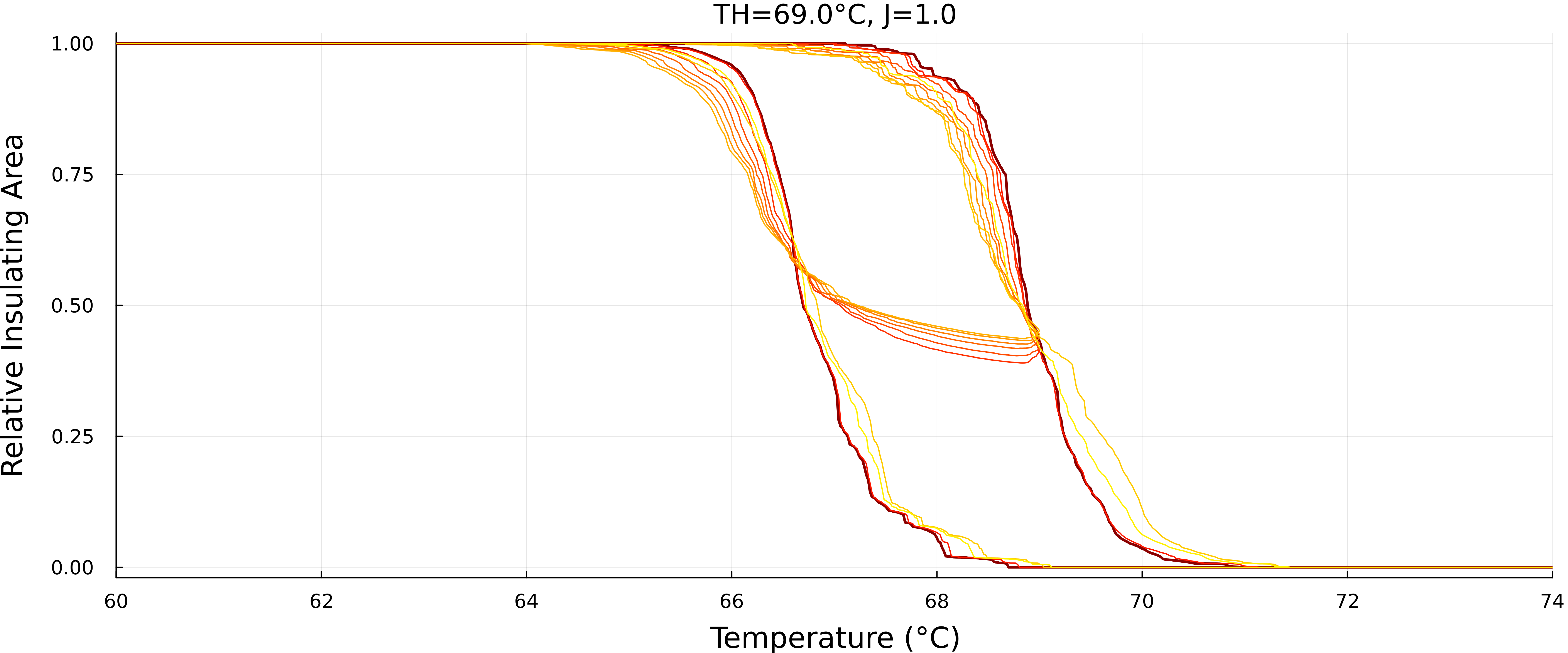}\\\includegraphics[width=0.98\textwidth]{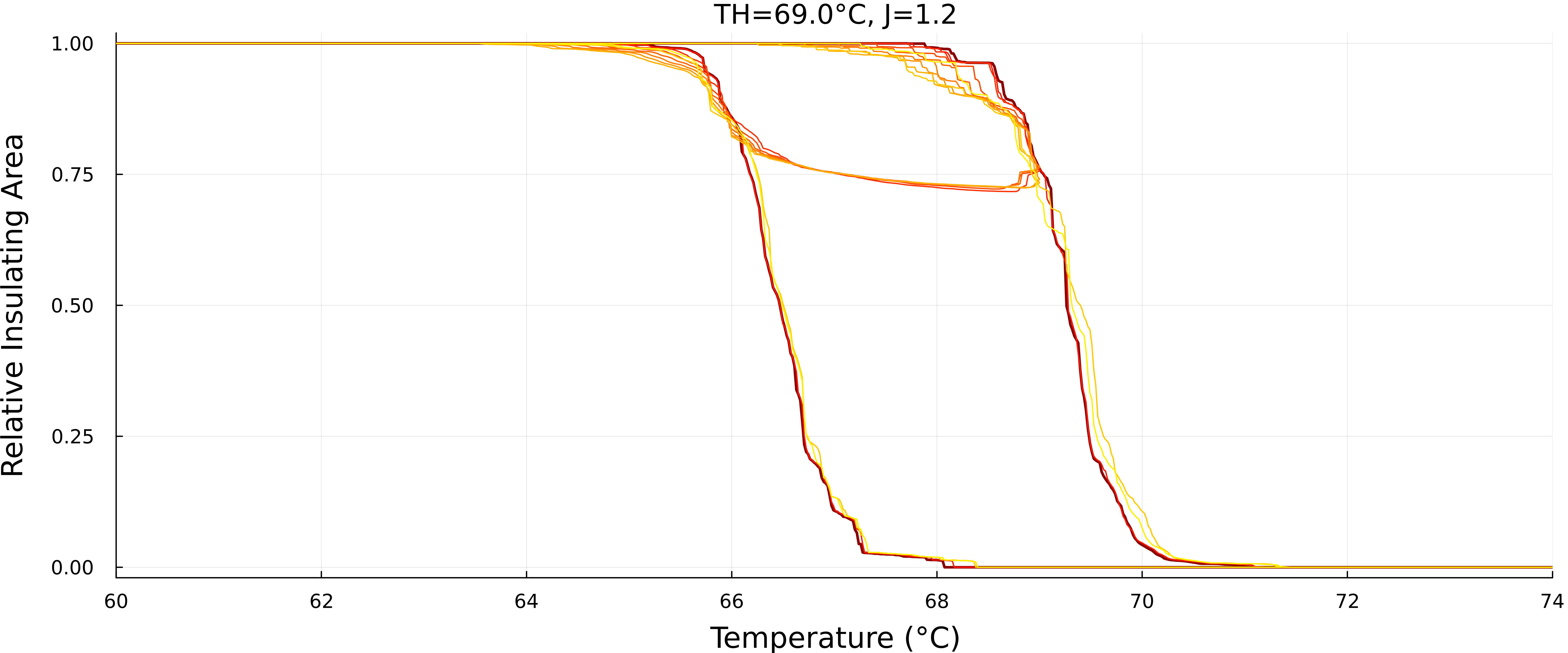}\\
    \includegraphics[width=0.98\textwidth]{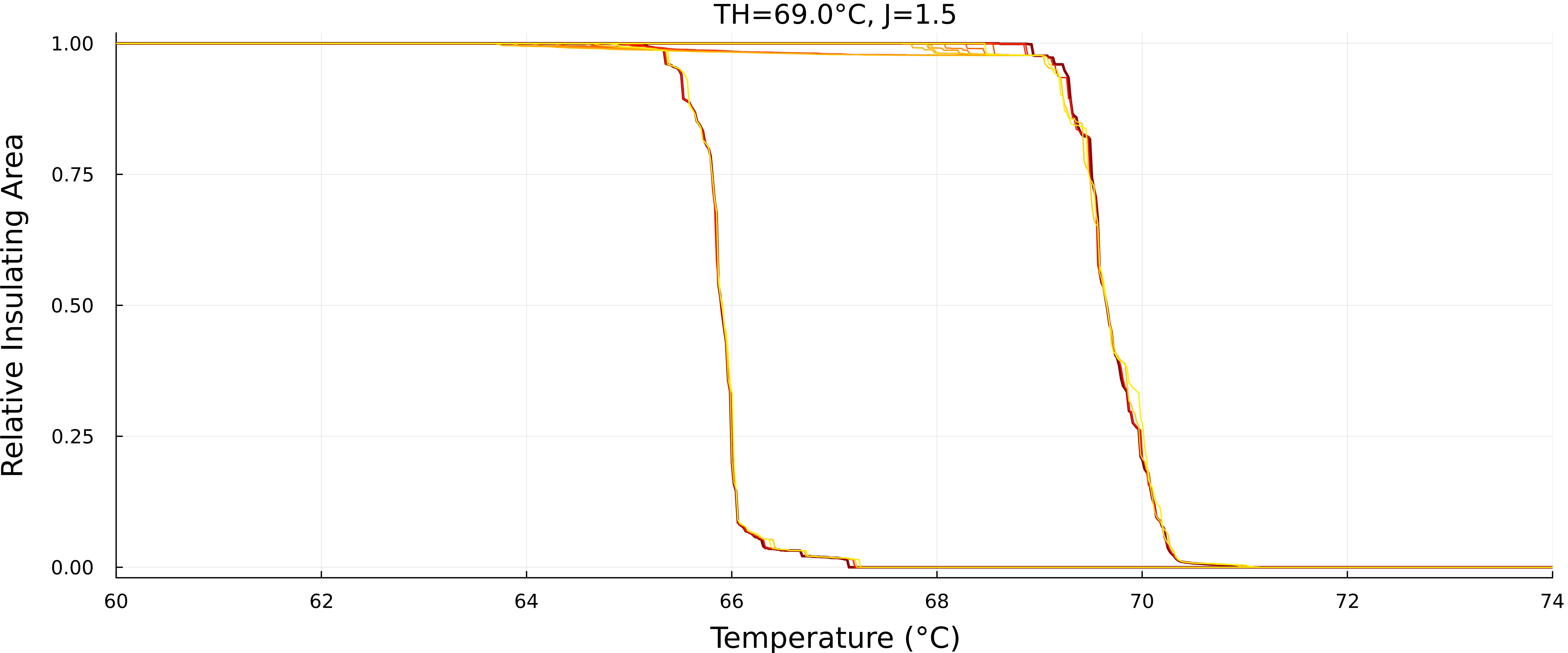}
    \caption{Simulated hysteresis loops with different interaction strength \(J\).}
    \label{fig:SI-loops}
\end{figure}

More \(J\) results are shown in Figure~\ref{fig:SI-loops}.

Choice of model parameters are kept the same as our earlier work~\cite{basakSpatiallyDistributedRamp2023}: 
For data with \( \num{900} \times \num{750} \) pixels, we assume 
\( l_x = \SI{33.6}{\um} \), \( l_y = \SI{28}{\um} \) and set 
\( T_{\text{H}} = \SI{68}{\celsius} \), 
\( \alpha = \frac{\mathrm{d} T_c}{\mathrm{d} (\rho / \rho_0)} = \SI{25}{\celsius} \), 
\( s = \rho_{\text{eq}}^{\text{insulator}} / \rho_{\text{eq}}^{\text{metal}} = 0.8 \) and diffusion coefficients 
\( D_i = \SI{61}{\nano\meter\squared\per\second} \), \( D_m = \SI{150}{\nano\meter\squared\per\second} \). 
In the absence of specific information about the species of defect(s), we have chosen the values of the diffusion coefficients to best match the experimental results.

\textcolor{mRed}{
\section{Experimental Data}
}

\begin{figure}[htbp]
    \centering
    \includegraphics[width=0.6\textwidth]{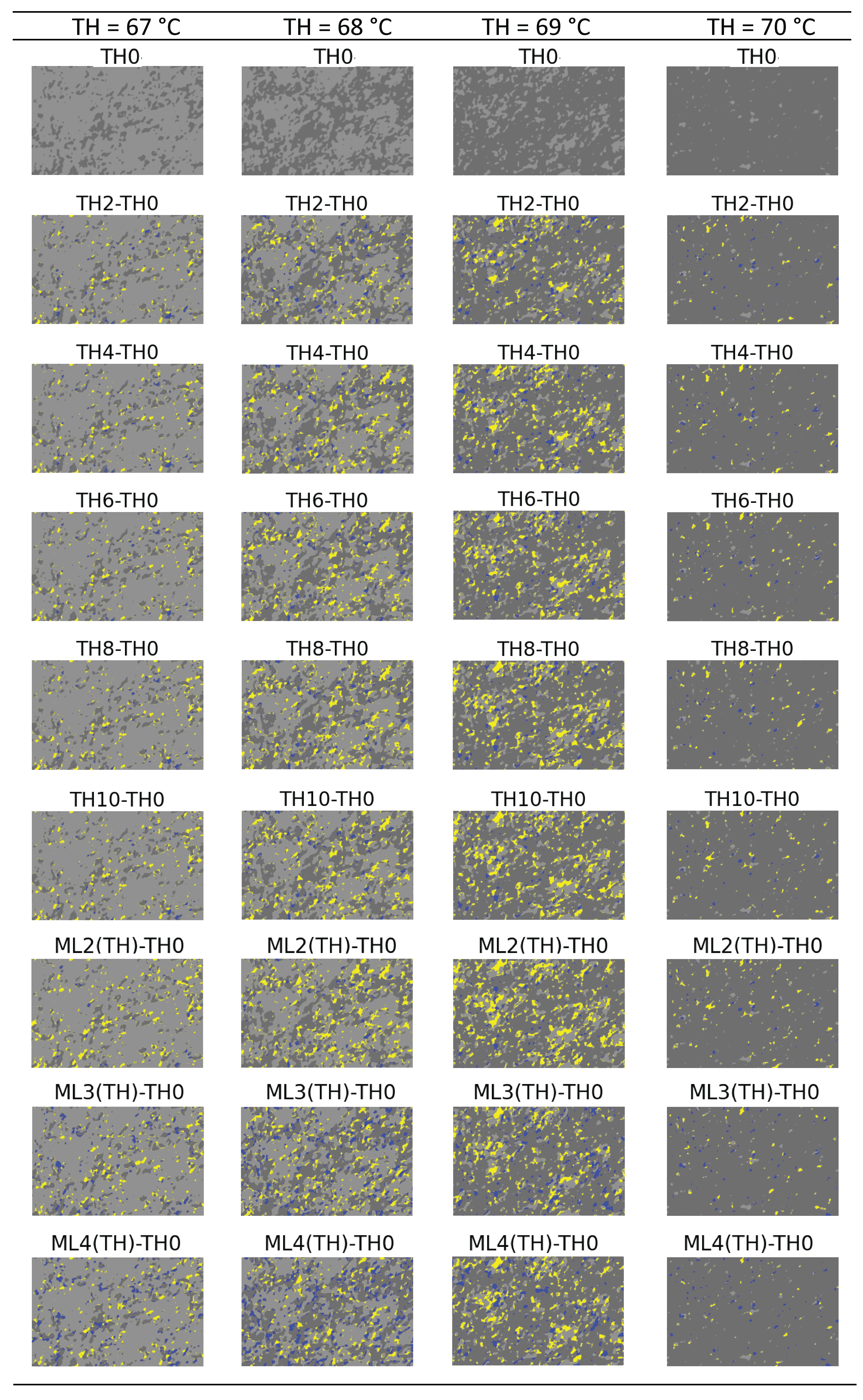}
    \caption{Color-coded VO$_2$ sample maps after each ramp reversal loop performed at various turning points \(T_\textnormal{H}\). Dark gray and light gray indicate unchanged metallic and insulating regions at \(T_\textnormal{H}\). Yellow regions represent metallic areas in the previous subloop that become insulating as subloops are performed. Blue regions indicate the opposite transition. Although yellow (insulating) and blue (metallic) regions are both created, insulating regions are predominant, making the sample more resistive macroscopically over time. 
    Changes in the fractal pattern on the sample as loops are performed constitute the memory training, called Ramp Reversal Memory, reported initially for a given \(T_\textnormal{H}\)
    ~\cite{basakSpatiallyDistributedRamp2023}. 
    Here, the maximum encoded memory is for \( T_\textnormal{H} = \SI{69}{\celsius} \), as reported in the main text, Figure~7(b). Sample images are all \SI{28}{\um} high.
    }
    \label{fig:MemoryMaps}
\end{figure}

\textcolor{mRed}{
Figure~\ref{fig:MemoryMaps} presents color-coded VO$_2$ sample maps showing spatially distributed memory accumulation in subloops. Yellow regions indicate metallic-to-insulating transitions while blue regions show the reverse.
}

\begin{figure}[htb]
    \centering
    \includegraphics[width=0.9\linewidth]{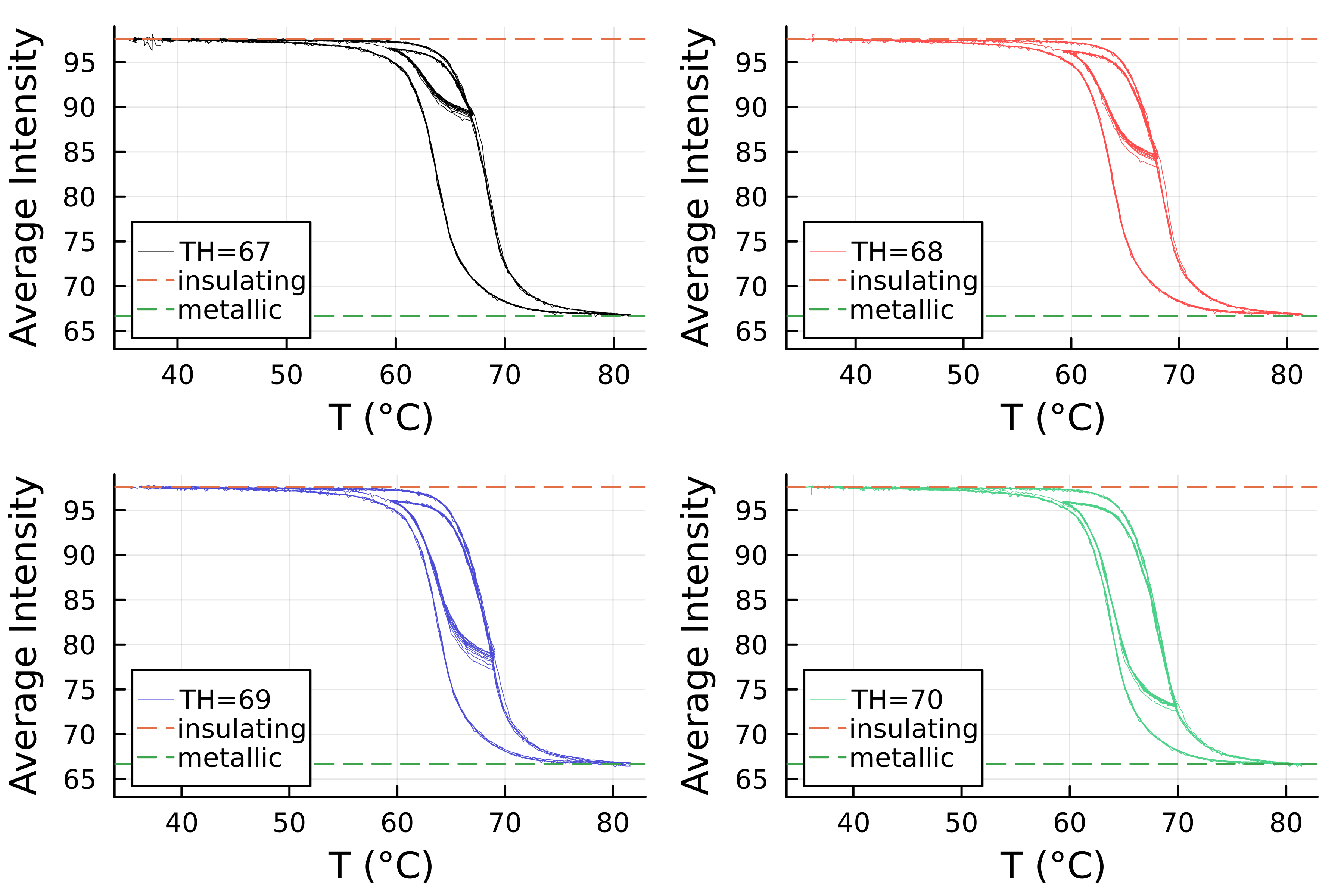}
    \caption{\textcolor{mRed}{Experimental hysteresis loops measuring average grayscale intensity during the temperature protocol of Figure~4 in the main text. The two horizontal lines correspond to average intensity \(I=97.6\) and
    \(I=66.7\) on a scale from 0 (black) to 255 (white).}}
    \label{fig:SI-exp_loops}
\end{figure}

\begin{figure}[htb]
    \centering
    \includegraphics[width=0.9\linewidth]{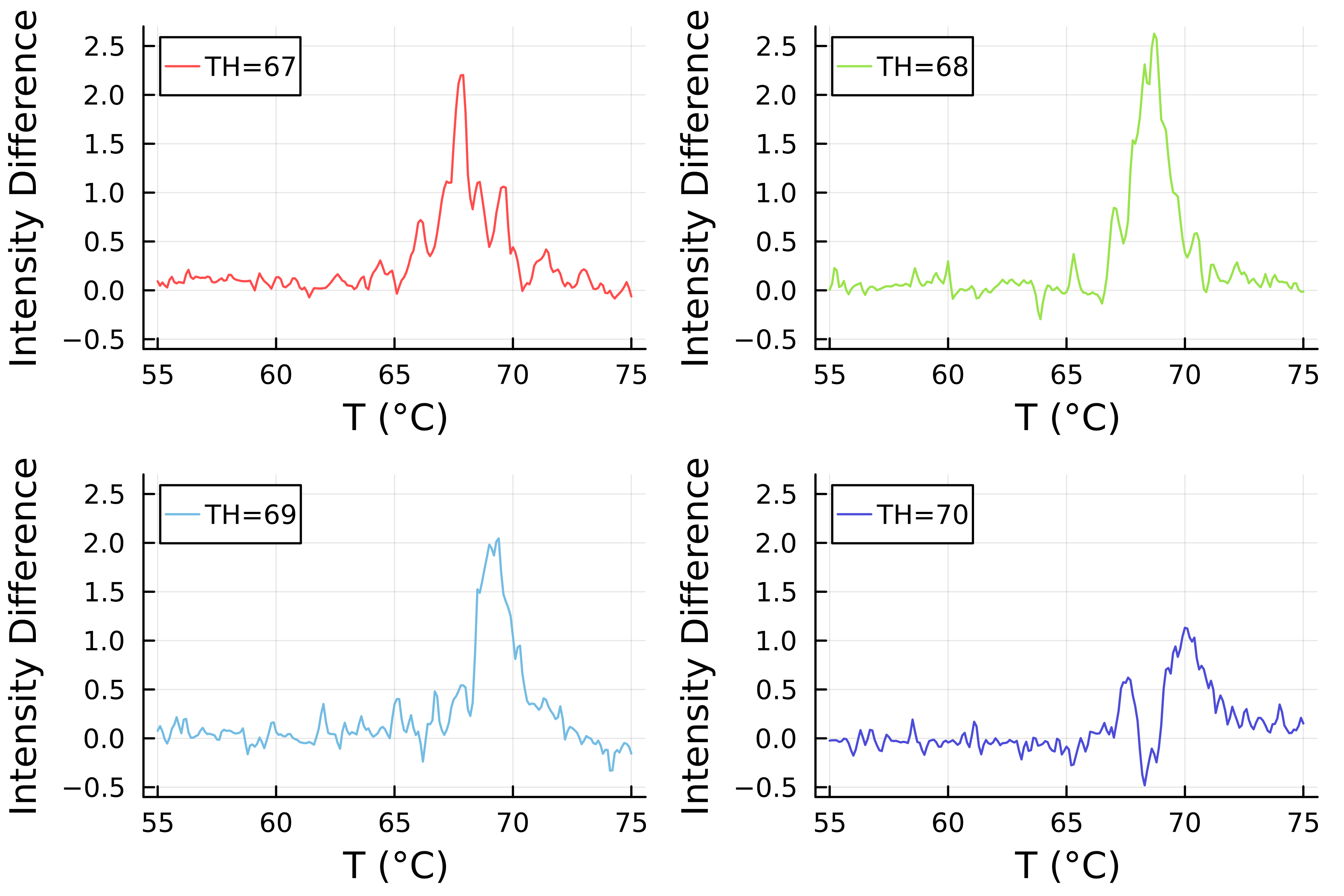}
    \caption{\textcolor{mRed}{Experimental data showing \(\Delta I\), the intensity differences between major loops measured immediately before and after the minor loops.}
    }
    \label{fig:SI-ML_diffs}
\end{figure}

\textcolor{mRed}{
Figure~\ref{fig:SI-exp_loops} shows experimental hysteresis loops used for data processing. The average intensity \(I\) corresponds to the optical measurement of reflected light using a CCD camera. Intensity is recorded on a grayscale from 0 to 255.
Figure~\ref{fig:SI-ML_diffs} plots intensity differences \(\Delta I\) between major loops for each \(T_\textnormal{H}\).
} 



\vspace{2cm}

\clearpage
\section{Reference}

\bibliography{SI_reference}